\documentclass[%
 aip,
 amsmath,amssymb,
 reprint,%
]{revtex4-1}
\usepackage{float}
\usepackage{graphicx}
\usepackage{dcolumn}
\usepackage{bm}

\usepackage[utf8]{inputenc}
\usepackage[T1]{fontenc}
\usepackage{mathptmx}
\usepackage{subfigure}
\begin{document}

\preprint{AIP/123-QED}

\title[]{Pulsed RF Schemes for Tearing Mode Stabilization}

\author{S. Jin}
 \email{sjin@pppl.gov}

\author{N. J. Fisch}
 \email{fisch@pppl.gov}
 \author{A. H. Reiman}%
 \email{areiman@pppl.gov}

\affiliation{ Department of Astrophysical Sciences, Princeton University, Princeton, New Jersey 08543, USA}%
\affiliation{Princeton Plasma Physics Laboratory, Princeton, New Jersey 08540, USA}%

\date{\today}

\begin{abstract}
The RF stabilization of tearing modes with current condensation has the potential to increase stabilization efficiency and loosen power localization requirements. Such benefits stem from the cooperative feedback between the RF deposition and resulting island temperature perturbation governed by diffusion. A self consistent treatment of the damping of an rf ray as it traverses the island shows that low damping scenarios can require unfavorably high powers to overcome initial power leakage and effectively capitalize on the nonlinear effect. In this work it is demonstrated that for such regimes, modulated stabilization schemes can achieve significant improvements in heating and current drive contributions to stabilization for the same average power as a continuous wave scheme. The impact of modulation frequency and duty cycle on the performance is explored, the results of which suggest modulation strategies in which the pulsing periods are kept on the order of a diffusive time. 
\end{abstract}

\maketitle

\section{Introduction}

Neoclassical tearing modes (NTMs) have been identified as one of the dominant causes of disruptions\cite{de_Vries_2011, de_Vries_2014}, and are anticipated to set the primary performance limit in ITER.\cite{ La_Haye_2006a,La_Haye_2009} Stabilization via current drive by rf waves\cite{Reiman_1983} has emerged as the leading solution, and has been the subject of much theoretical \cite{Zohm_1997,Yu_1998,Harvey_2001,Bernabei_1998,Kamendje_2005,La_Haye_2006b,La_Haye_2008,Sauter_2010,Smolyakov_2013,Ayten_2014,Borgogno_2014,F_vrier_2016,Li_2017,grasso_borgogno_comisso_lazzaro_2018,Grasso_2016} and experimental work\cite{Zohm_1999,Prater_2003,Warrick_2000,Gantenbein_2000,Zohm_2001,Isayama_2000,La_Haye_2002,Petty_2004,Volpe_2015,Nelson_2019}. Prior to the treatment in Ref. \onlinecite{Reiman_2018}, stabilization calculations were done without self-consistently considering the effect of wave deposition on island temperature. This traditional approach neglects strong nonlinear effects and the opportunities to exploit them. 

The most studied waves for rf stabilization are the lower hybrid wave which drives currents through the LHCD effect \cite{fisch_1978} and the electron cyclotron wave through the ECCD effect\cite{fisch_1980}. Although other rf waves might be enlisted to drive current \cite{fisch_1987}, these waves are highly sensitive to changes in temperature: $P_{dep}\propto n_{res} \propto exp(-w^2)$ where $P_{dep}$ is the local power deposition, $n_{res}$ is the number of resonant particles and  $w=v_{ph}/v_{th}$ is the ratio of the phase and thermal velocities.\cite{fisch_1987,fisch_1980} A small temperature perturbation $\widetilde T$ then contributes an exponential enhancement factor $exp(u)$,  where $u:=w_0^2\widetilde T/T_0$, $T_0$ is the unpertubed temperature, and $w_0^2\approx 4-20$. The exponential dependence on $u$ and typically large values of $w_0^2$ translate small temperature perturbations into large effects on deposition. The thermal insulation provided by the closed magnetic topology of the island can lead to significant temperature perturbations \cite{Westerhof_2007} governed by perpendicular diffusion. In combination, this amounts to a significant nonlinear enhancement of power deposited and corresponding driven current. Additionally, an initially broad deposition profile can be effectively narrowed due to the centrally peaked temperature profiles. This amplification and focusing are termed the current condensation effect, predicted in Ref. \onlinecite{Reiman_2018}, and suggest that traditional calculations of efficiency and localization requirements have been underselling  rf stabilization. 

Subsequent work self consistently treating the damping of an rf ray as it traverses and heats the island \cite{Rodriguez_2019} revealed two potential concerns for stabilization scenarios, hereafter referred to as \textit{leakage} and \textit{shadowing}. The first refers to poor absorption of the rf when the deposition width is large compared to the island width; such scenarios suffer the disadvantage of requiring high powers to effectively utilize the input energy and capitalize on the nonlinear effect. While the latter is primarily a concern in strong damping scenarios not discussed here, high enough powers can enhance deposition at the island periphery at the cost of the center for any damping strength, negatively impacting stabilization efforts. 

It will be shown here that modulated schemes can unlock heating and stabilization enhancements for the same cycle averaged power as their continuous wave counterparts, when the deposition width is comparable to or larger than the island width. The effect requires a sufficiently high peak power to overcome leakage with the nonlinear effect, and is optimized for pulse periods on the order of a diffusion time which avoids shadowing. The rf capabilities of present devices are estimated to meet both the power and modulation frequency requirements for accessing this effect\cite{Rodriguez_2019}. The requirements on relative island and deposition widths lend these results particular relevance to LHCD for steady state scenarios, where a broad deposition profile may focus within the island due to the current condensation effect. Pulsing on diffusive timescales can then further exploit the nonlinear effect to cut average power costs and minimize peripheral deposition. 

The paper is organized as follows: Section II introduces the time-dependent coupled rf-island model, summarizes key features of steady state solutions motivating pulsed schemes, the time dependence of which are then explored in detail. Section III presents the heating and stabilization properties of pulsed schemes, in particular how performance may be optimized by choice of the modulation parameters, and discusses accessibility constraints. The implications of the results for developing rf stabilization strategies and their present experimental relevance are discussed in Section IV. Finally, the main conclusions are summarized in Section V. 

\section{Coupled Wave Damping and Island Heating Model}
In order to develop an intuition for the benefits of pulsed schemes, the simplest possible model that captures the physics of wave damping and island heating will be used.
On the time scales of interest, the electron temperature evolution of the island interior is described by the heat diffusion equation:
\begin{equation}
    \frac{3}{2} n \partial_t T -\nabla\cdot(\kappa_\cdot \nabla T)=P
\end{equation}
where n is the plasma density, $\kappa$ is the heat conductivity tensor, and $P$ is the volumetric power deposited by the rf wave, the form of which will be specified shortly. This equation can be simplified by adopting a 1-D slab model for the island geometry, treating flux surfaces isothermally, and linearizing for small perturbations of the island temperature $\widetilde T$ relative to the surrounding plasma temperature $T_0$:
\begin{equation}
     \frac{3}{2} n \partial_t T -\kappa_\perp \partial_x^2 T=P
\end{equation}
subject to the boundary conditions $\widetilde T (x=\pm W_i/2) $, where $W_i$ is the island width. A detailed treatment of the boundary conditions and time scale orderings used to make these simplifications can be found in Ref. \onlinecite{Rodriguez_2019}, with the difference here being that the present model retains time dependence on diffusive time scales--therefore the electron temperature is considered alone, neglecting the slower process of heat loss to ions. The details of this time scale ordering and its consequences can be found in Appendix B.

The power source that enters the diffusion equation is related to the wave energy density V in the following way:
\begin{equation}
    P=-(V'(x)+V'(-x))/2
\end{equation}
which simply describes that the energy lost by the ray at some location goes into heating the plasma on the local flux surface. The symmetrized form results from the 1-D slab geometry, as the power deposited at a given location $x$ is shared with the whole flux surface, also labeled with $-x$. 

Express the ray's spatial damping as $ V'=-\alpha V$. The damping strength $\alpha$ is in general a complicated function of the wave and plasma parameters\cite{Bonoli_1986,Prater_2008}; for the purposes of studying how the wave damping couples to the island heating it can be simplified through the following reasoning. For waves acting on the tail of a Maxwellian distribution, the damping is proportional to the number of resonant particles at the wave's phase velocity, and exponentially sensitive to small temperature perturbations: 
\begin{equation}
    \alpha \propto n_{res} \propto \exp{(-m v_{ph}^2/2 T)} \approx \exp{(-w_0^2)} \exp{w_0^2 \widetilde T/ T_0}
\end{equation}
where $w_0^2 = m v_{ph}^2/2 T_0$. It is important to note here that this proportionality factor is typically large for the waves of interest ($\approx 10-20$ for LH and $\approx 4-10$ for EC), which means that small temperature perturbations can dramatically affect deposition. Therefore, this exponential enhancement factor has the strongest impact on damping strength. It can be written as an explicit factor $\alpha=\alpha_0 \exp{(u)}$ with the weaker dependencies suppressed in the linear damping strength $\alpha_0$.  

The wave damping equation then takes the form
\begin{equation}
    V'=-\alpha_0 V \exp{( w_0^2 \widetilde T/T_0)}
\end{equation}
with $V(x=-W_i/2)=V_0 (t)$, the input energy as the wave enters the island. The equations can be further simplified by adopting the normalized temperature $u:=w_0^2 \widetilde T/T_0$, and the following scalings for space, time, damping strength, and wave energy density respectively: $x_{scl}=W_i/2, \quad t_{scl}= 3 W_i^2/8 \chi_\perp, \quad \alpha_{0,scl}=2/W_i, \quad V_{scl}=2 n T_0 \chi_\perp/W_i w_0^2$. The quantity $t_{scl}$ here may also be identified as the electron diffusion time $t_{D,e}$. The final form of the coupled diffusion and wave damping equations is then: 
\begin{equation}
    \dot u - u''= -(V'(x)+V'(-x))/2 
\end{equation}
\begin{equation}
    V'=-\alpha_0 V \exp{(u)}
\end{equation}
subject to the boundary conditions (i) $u(x=\pm 1)=0 $ and (ii) $V(x=-1)=V_0 f(t)$. 
In order to expose the key effects of pulsing, we specialize to the case where the power damping is only regulated by the temperature, rather than by the magnetic field or other details of the wave trajectory. This means that for constant temperature, there is pure exponential damping and therefore the deposition will be highest at the island periphery where the power is not yet depleted. Centrally peaked deposition profiles can still be achieved with centrally peaked island temperature profiles granted by slow cross-field diffusion.

\subsection{Summary of steady state solutions}

\begin{figure}[b]
\centering     
\includegraphics[width=\linewidth]{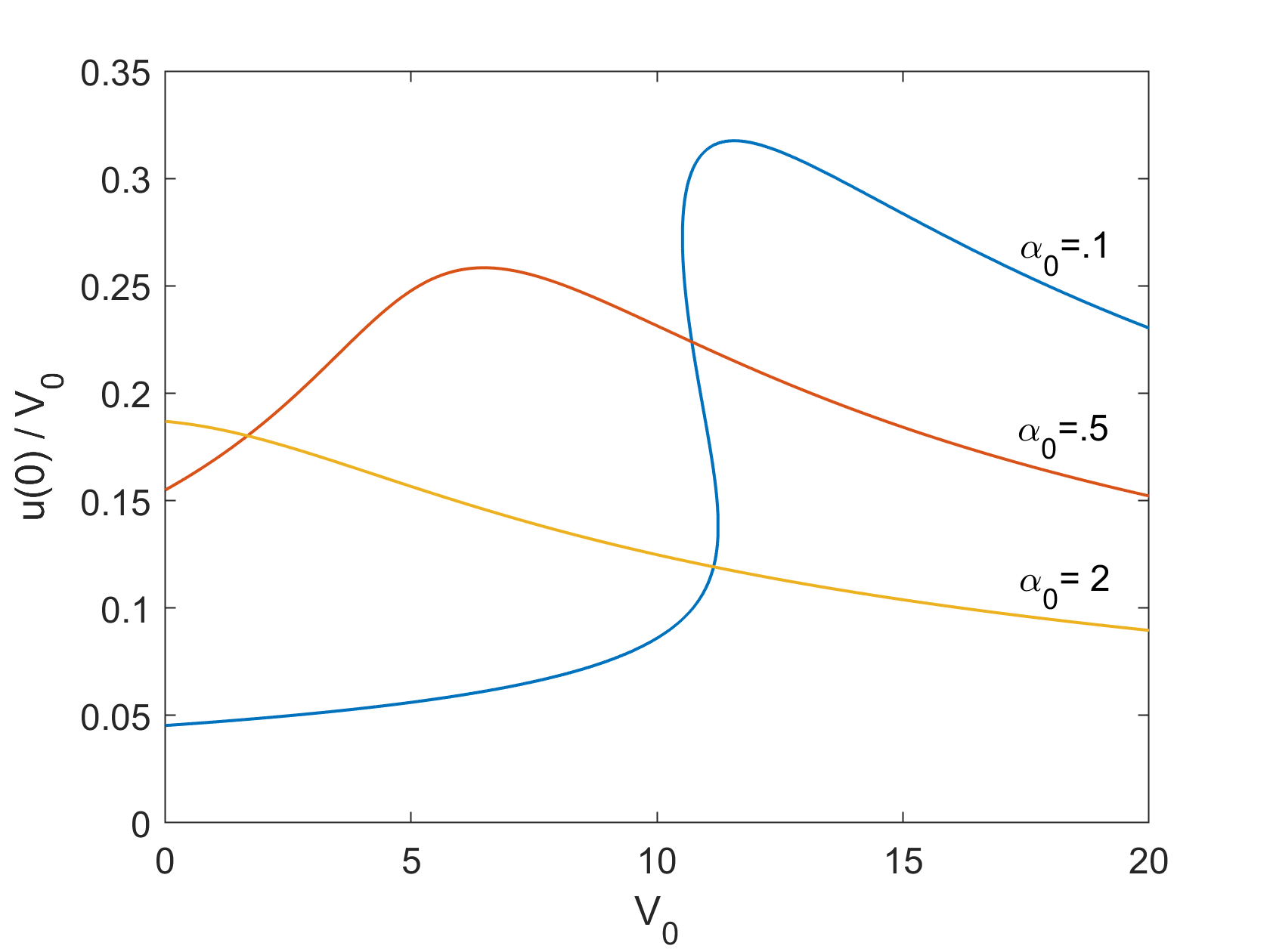}
\caption{\label{fig:usteady}Heating efficiency $u(0)/V_0$ vs $V_0$ at various fixed damping strength $\alpha_0$. It can be seen that for low damping regimes ($\alpha_0=0.1,0.5)$, low powers fail to achieve efficient heating. At high enough powers absorption improves, but further power increase deteriorates heating efficiency due to peripheral depletion. At high damping ($\alpha_0=2$), efficiency monotonically decreases with increasing power.}
\end{figure}
Previous work with this coupled wave-island system \cite{Rodriguez_2019} has been done in the steady state ($f(t)=1$, $\dot u \rightarrow 0$), in which case the system is fully characterized by the two parameters $\alpha_0 $ and $V_0$. The parameter $\alpha_0$ provides a natural separation of the parameter space into a low-damping ($\alpha_0 <\sim1$) and high-damping regime ($\alpha_0 >\sim1$), that are differentiated by the response of the heating efficiency $u(0)/V_0$ with power input $V_0$ as illustrated in Fig. \ref{fig:usteady}. 

The high damping regime is characterized by monotonically decreasing heating efficiency, as even linearly the power is effectively absorbed. Increasing $V_0$ immediately contributes to shadowing, with rising temperatures causing the rf to be depleted ever further in the periphery. In contrast, low damping regimes have initially poor heating efficiencies due to significant energy leakage ($1-V(1)/V(-1) << 1$), i.e. most of the input energy simply passes through without being absorbed by the island. With higher $V_0$ and correspondingly larger island temperatures, the enhanced damping improves heating efficiency until it reaches a maximum once the rf is effectively absorbed ($V(1)/V(-1) <<1$). 

At very low damping ($\alpha_0 <\sim .2$) this transition occurs suddenly, with a narrow band of powers just below this threshold where two stable solutions exist, the hotter solution corresponding to low leakage. This solution structure allows for a hysteresis effect in that after jumping to the hot solution branch, it is possible to then reduce power and stay on the hotter branch, as discussed in detail in Ref. \onlinecite{Rodriguez_2019}. It is important to note that the pulsed enhancements explored in this work exist for a much broader range of $\alpha_0$ and $V_0$, although the physics is similar in that high powers are used to gain access to a regime with efficient absorption despite low linear damping. Increasing $V_0$ past this point eventually leads to diminishing gains in temperature ($u(0)\sim \text{log} (V_0/2\alpha_0+1), \text{ as } V_0 \rightarrow \infty$), as shadowing takes over.

The remainder of this paper will be confined to the low damping regime, where pulsed schemes have the opportunity to exploit the nonlinear improvement of heating efficiency with $V_0$ to obtain significantly improved cycle-averaged temperatures for the same average power. The eventual shadowing at high powers is relevant to understanding the optimization of pulse times, as will be elaborated in the time dependent picture.

\subsection{Time evolution of heating pulse}
 
\begin{figure}[b]
\centering     
\includegraphics[width=\linewidth]{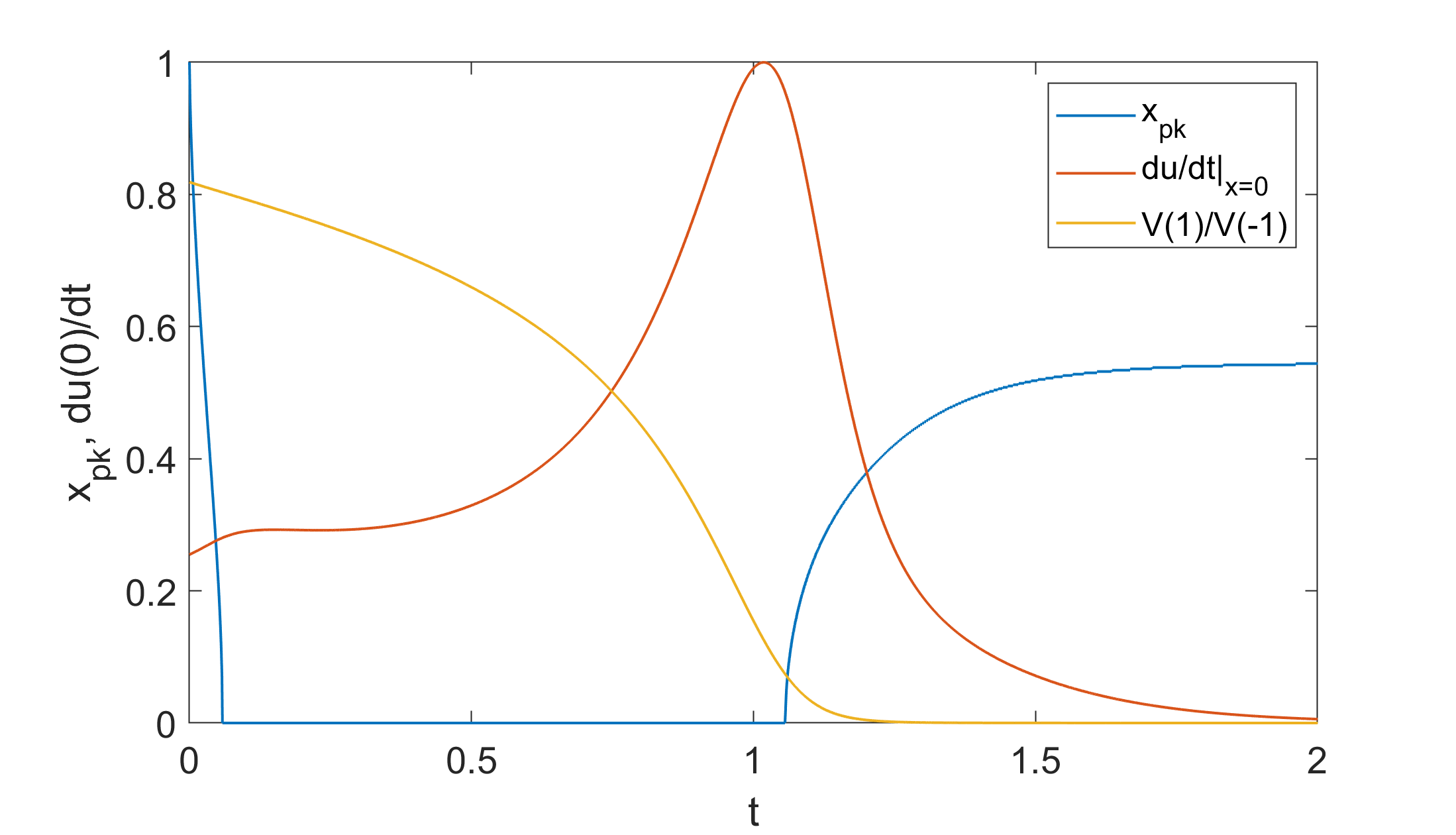}
\caption{\label{fig:helpful}Blue: $x_{pk}$, Orange: $\frac{du}{dt}|_{x=0}$, Yellow: Leakage (V(1)/V(-1)) Evolution of location of peak deposition, growth of central island temperature, and power leakage upon the application of heating for the $\alpha_0=.1$, $V_0=10$ case.}
\end{figure}

In this work, time dependence is introduced through periodic $f(t)$, such that a ``quasi steady state" (in the sense that system quantities (u,V) averaged over a power cycle are constant in time) is achieved. Furthermore, $f(t)$ will always be taken to be of the form 
\begin{equation}
   f(t)=%
   \begin{cases}
     1 & \text{if $ t \text{ mod } \tau<\tau_{on}$} \\
     0 &\text{otherwise}
   \end{cases}
\end{equation} 
where $\tau$ is the period of pulsing. Therefore, the island-wave system is characterized by $(\alpha_0,V_{0,eff},\tau,d)$, where $\tau$ is the period of pulsing, $V_{0,eff}$ and $ V_{0}$ are the cycle averaged and instantaneous powers respectively, and $d=\tau_{on}/\tau=V_{0,eff}/V_{0}$ is the duty cycle. $\alpha_0$ remains entirely unchanged from the steady state model, the cycle averaged $V_{0,eff}$ corresponds to the steady state $V_0$, while $\tau$ and $d$ are new degrees of freedom introduced by the pulsing.

In order to understand the impact of these modulation parameters on the performance of pulsed schemes, it is instructive to dissect the time evolution of the island-wave system as it approaches a steady state (Fig. \ref{fig:helpful}). 
 It can be seen that due to the linearly low damping and resulting initially broad deposition, diffusive edge losses are able to quickly produce a centrally peaked temperature profile that pulls the location of maximum deposition $x_{pk}$ to 0. Central heating then rapidly accelerates as the absorption improves. This transient phase of the heating process exhibits highly favorable heating and stabilization properties, suggesting that an optimized pulsed scheme would require a $\tau_{on}$ long enough to capture its full duration.

 Now examining the latter half of the heating process, as the leakage saturates to 0, the central temperature stops growing and $x_{pk}$ moves back out of the center--this transition event will be termed the shadowing onset time. This shadowing occurs as the enhanced damping due to the rising island temperatures cause the incoming wave to be depleted ever further in in the periphery. Therefore, once the island is heated long enough for power to be efficiently absorbed, further heating amounts to diminishing gains in temperature and can actually reduce the current driven at the O-point. It follows that an optimum pulse time would be roughly the shadowing onset time, but shorter due to residual temperature from the previous pulse.  Therefore, the optimum pulse time is set by heating with a sufficiently high instantaneous $V_{0}$, long enough to overcome leakage, but not so long that unfavorable shadowed deposition occurs. The off time between pulses plays a more indirect role through setting $V_0$ for a given time averaged power and by determining how much residual temperature there is at the start of the next cycle. 
 
\section{Pulsed Stabilization Schemes}
\subsection{Optimizing performance with pulse frequency}

The merit of a pulsed scheme can be judged from the resulting temperature and current profiles. Large temperature enhancements can be understood to be favorable even in the absence of rf current drive, from the resulting perturbations to the Spitzer conductivity (from $\sigma_{Sp} \propto T_e^{(-3/2)}$).\cite{Kurita_1994,Hegna_1997,De_Lazzari_2009} The benefits only get more dramatic once considering the exponential enhancement factor carried by the driven current ($j_{CD}\propto P_{dep} \propto exp(u)$). Stabilization efforts therefore benefit from centrally peaked, large amplitude temperature profiles, qualities which are reflected in the summary measure of heating efficiency $u(0)/V_0$. This metric suffers from either poor absorption or broad temperature profiles resulting from shadowed deposition.

The stabilizing power of the deposition profiles can also be captured in the metric $P_{cent}=\int _{-.5}^{.5} P dx/ V_0$, which gives the fraction of power deposited in the center half of the island to the total power that is available to the island. While a more direct calculation based on the modified Rutherford equation as described in Ref. \onlinecite{eduardo} is also possible, $P_{cent}$ provides a more sensitive metric for studying the impact of modulation parameters on performance, and is more useful for this work. Further discussion of how the figures of merit used here compare to the traditional stability metric can be found in Appendix A.

\begin{figure}[t]
\centering     
\includegraphics[width=\linewidth]{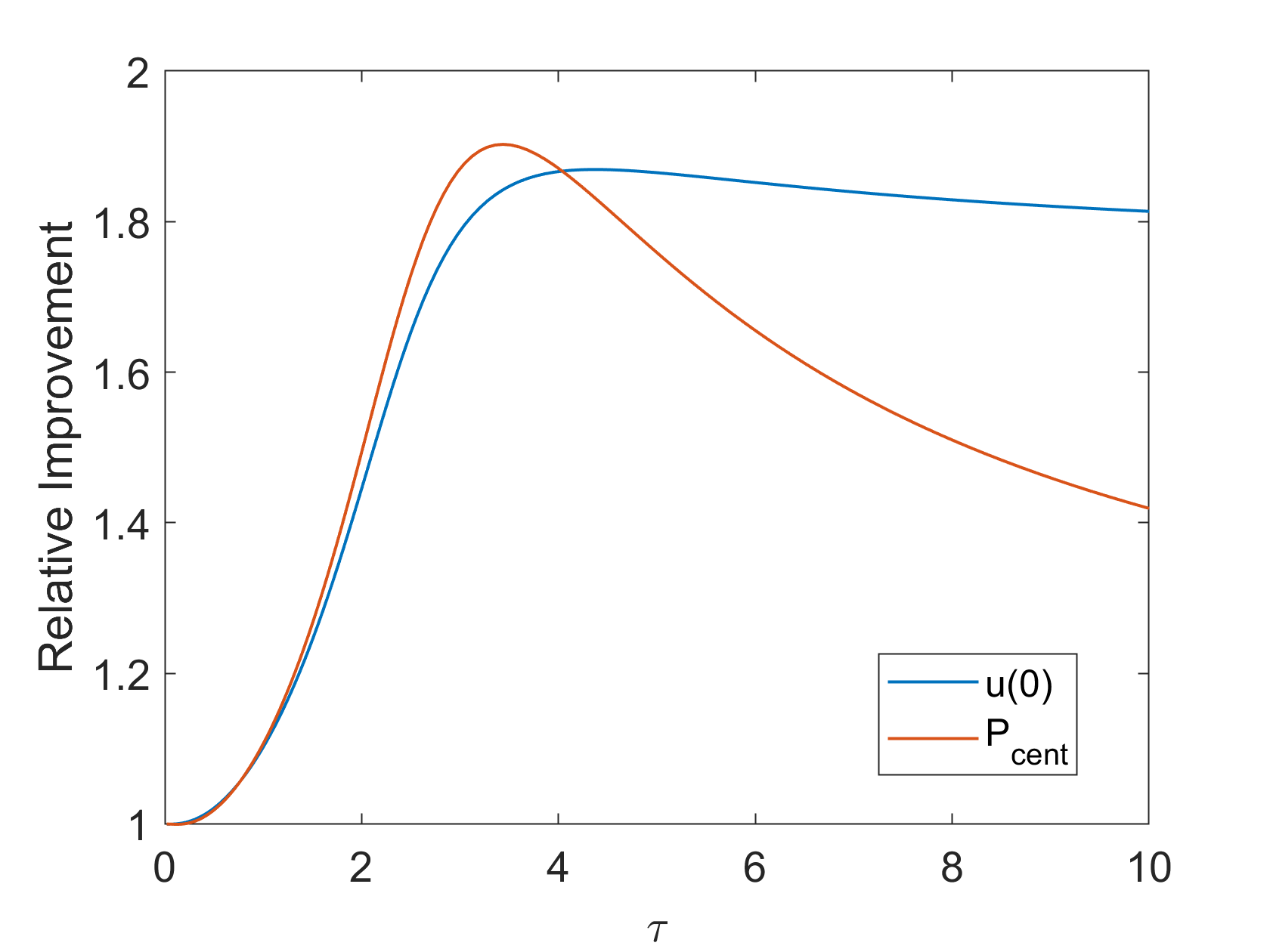}
\caption{\label{fig:tauscan}Demonstration of stabilization scheme performance dependence on pulse times, all quantities normalized to the equivalent steady state for the ($\alpha_0=.1, V_0=4$) case }
\end{figure}

\begin{figure}[b]
\centering     
\includegraphics[width=\linewidth]{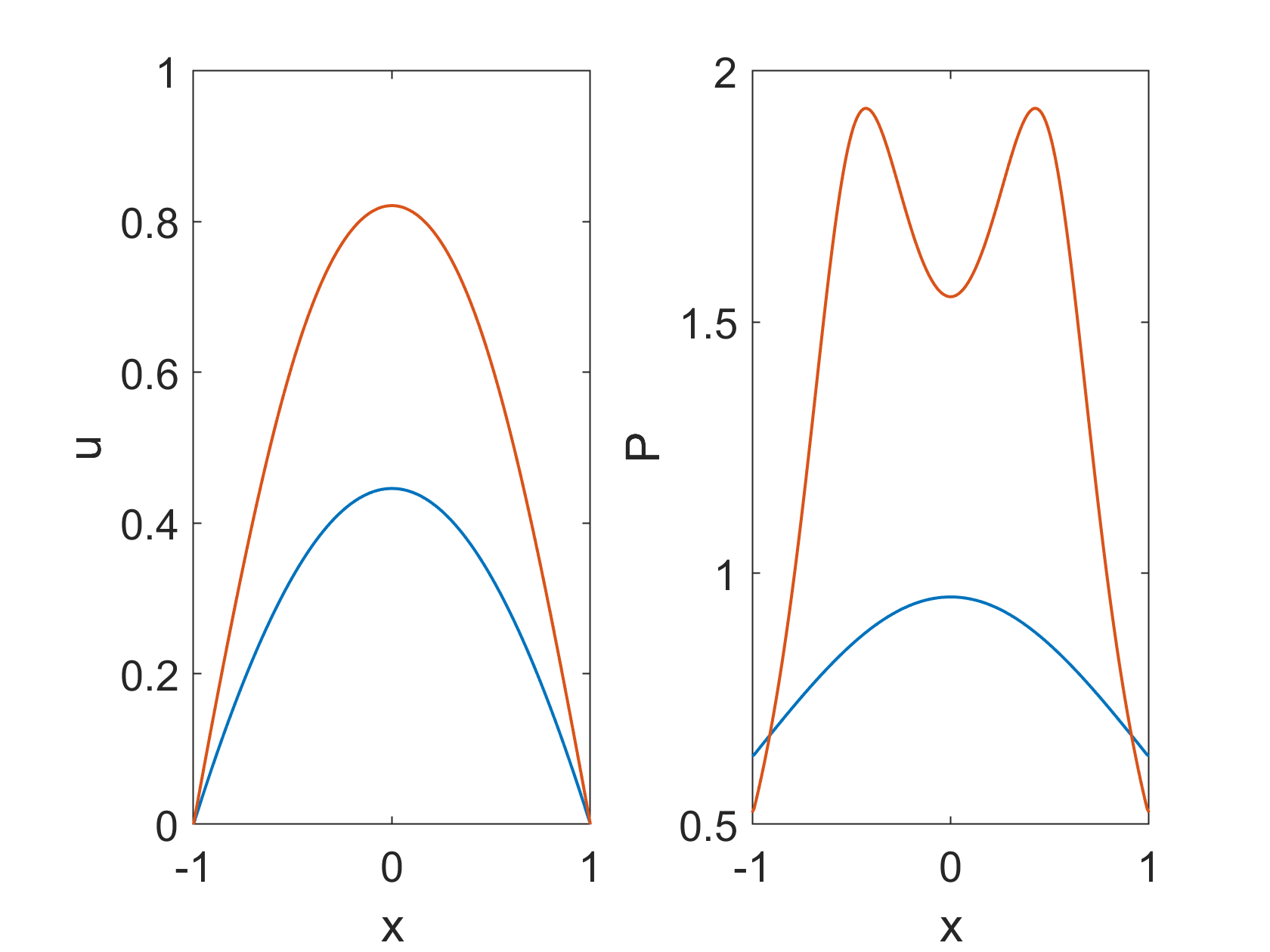}
\caption{\label{fig:upcomp}Comparison of temperature (left) and current (right) profiles achieved through optimally pulsed (orange) and continuous wave (blue) schemes for $\alpha_0=.1$ and $V_0=4$}
\end{figure}

The reasoning developed in the previous section anticipates an optimum pulse period on the order of a diffusive time, in order to reap the benefits of the transient heating period but avoid the shadowed steady state behavior. Figure \ref{fig:tauscan} demonstrates this for the ($\alpha_0=.2$, $V_0=4$, $d=.2$) case. Intuitively, the $\tau \rightarrow 0$ limit corresponds to the steady state, as diffusion has no time to act in between pulses, so the model here smooths out to the steady state solution for $V_{0,eff}$. As $\tau>>1$ the system approaches the steady state solution for $V_{0}$ weighted by the duty cycle, but in this limit, other physics not included in this model will need to be accounted for. Both figures of merit improve with increasing $\tau$ until reaching an optimum around $\tau \approx 3$, corresponding to a heating time $\tau_{on}\approx 0.6$. Evidently the central temperature $u(0)$ and central power deposition $P_{cent}$ improve rapidly with increasing pulse time, as the system is given more time to be in the favorable central heating stage (early portion of Fig. \ref{fig:helpful}). As the pulse times are further increased, and the system is allowed more time in the shadowed saturation period (latter portion of Fig \ref{fig:helpful}), $P_{cent}$ strongly reflects this suboptimality while the central temperature $u(0)$ is not as dramatically affected. Nevertheless, as both metrics are roughly simultaneously maximized, it makes sense to speak of a unique optimal pulsing time $\tau_{opt}$. Figure \ref{fig:upcomp} shows the optimally pulsed cycle-averaged temperature and current profiles compared to their continuous wave counterparts.


\begin{figure}[h]
\centering     
\includegraphics[width=\linewidth]{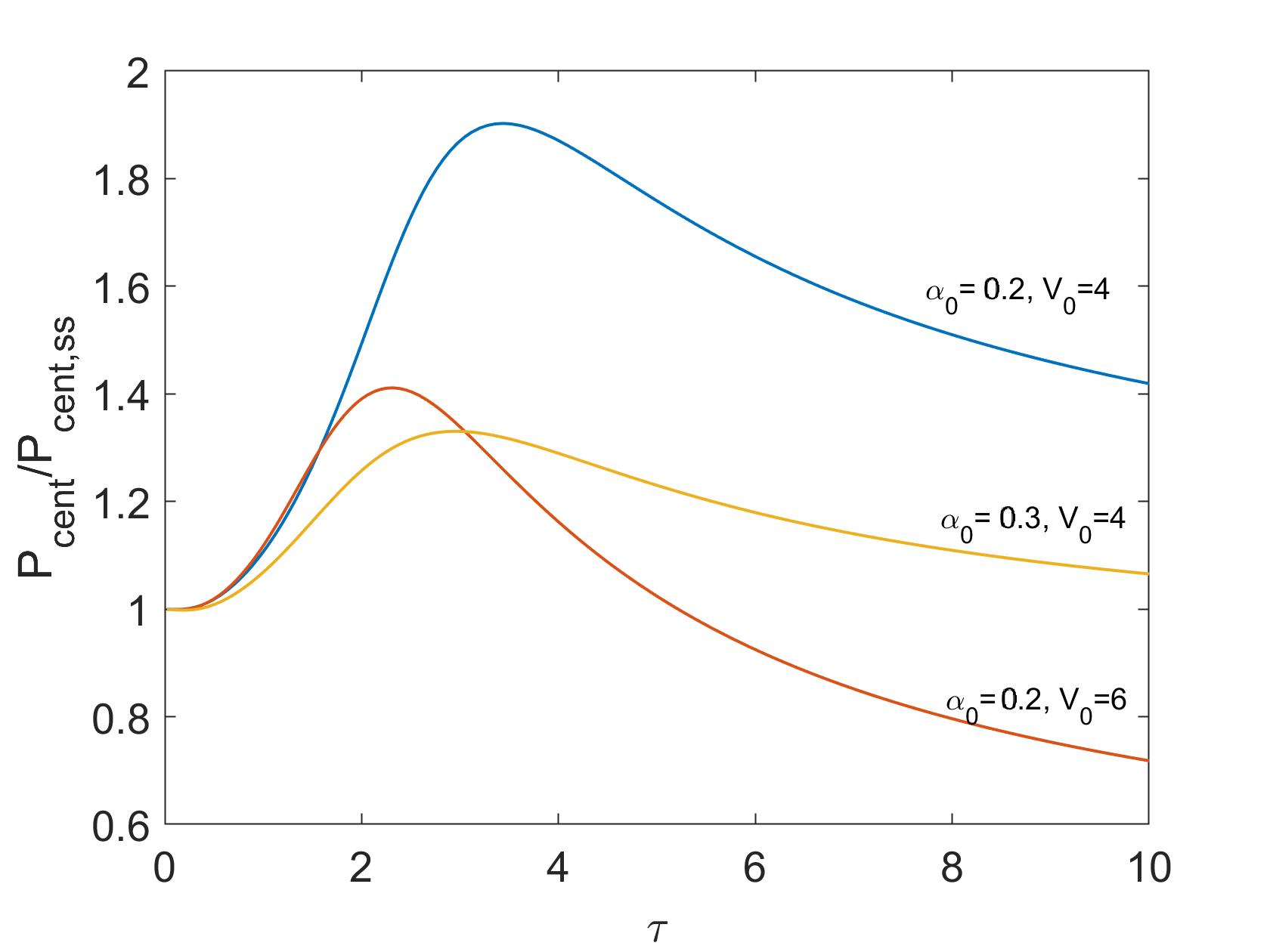}
\caption{\label{fig:paramcomp}Comparison of relative $P_{cent}$ improvements for 3 representative cases, all performed at duty cycle $d=.25$.}
\end{figure}

As either damping strength $\alpha$ or power $V_0$ is increased (at fixed duty cycle $d$), the optimal pulsing time decreases as seen in Fig. \ref{fig:paramcomp}. This is due to higher $\alpha_0$ corresponding to better linear absorption, so shorter nonlinear heating time necessary to overcome leakage. Higher powers similarly require less heating time to accomplish the amount of edge temperature increase for the onset of shadowing. There is also less to be gained from optimally pulsing for either comparison case (at higher $\alpha_0$ or $V_0$), which can be understood in the same framework of using pulsing to overcome leakage. If leakage is less of a concern to begin with, as would be the case for higher $\alpha_0$ or $V_0$, there is simply not as much room for improvement. 
It can also be seen that pulsing only negatively impacts performance relative to the steady state at pulse times much longer than diffusion times. This is due to the shadowed deposition associated with higher powers occupying greater portions of the heating pulse.
\subsection{Accessibility Caveats}
\begin{figure}[h]
\centering     
\includegraphics[width=\linewidth]{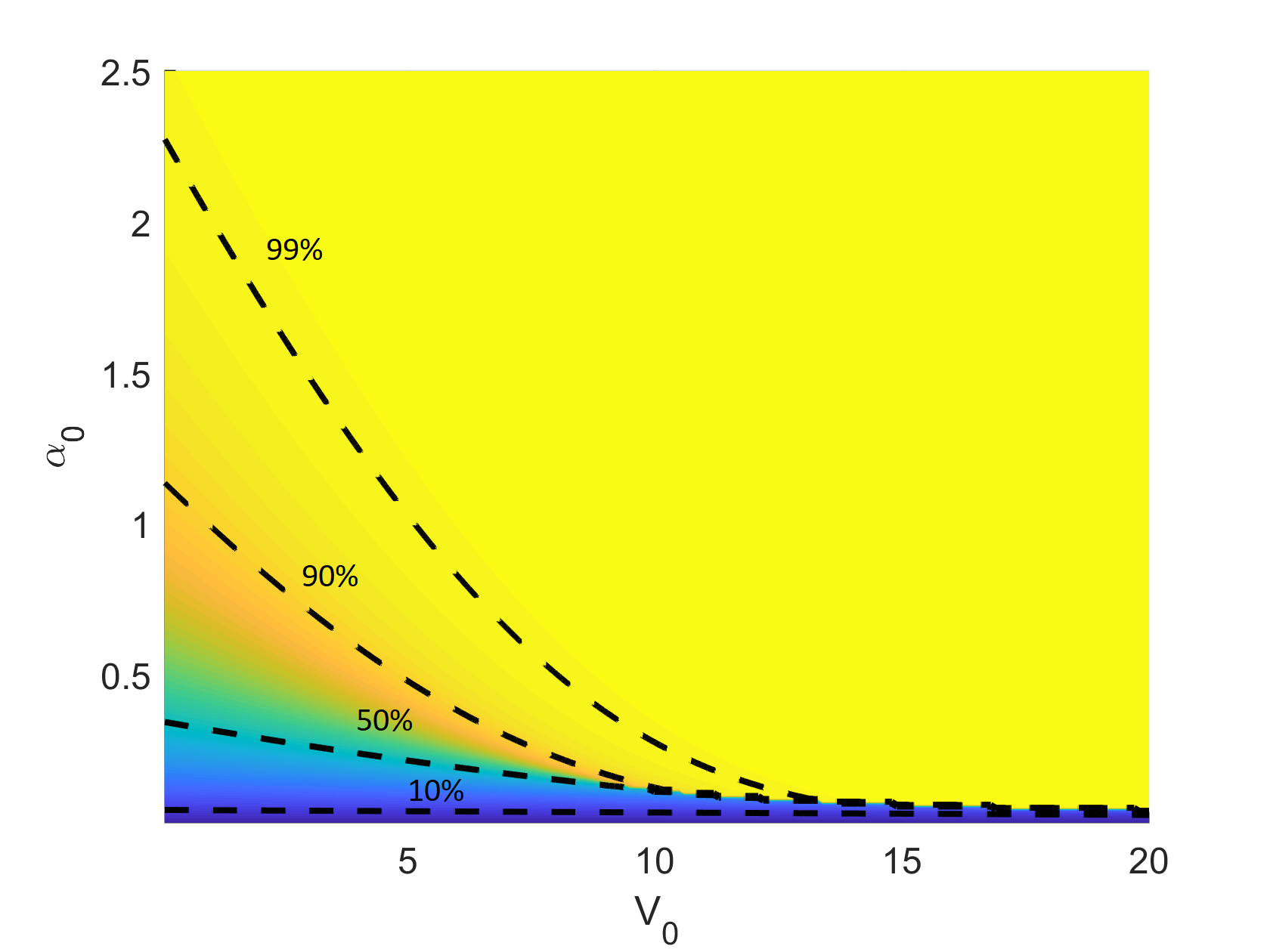}
\caption{\label{fig:leak}Contour map of absorption $1-V(1)/V(-1)$ in the $\alpha_0, V_0$ steady state parameter space}
\end{figure}

As suggested by the weaker improvements for increased $\alpha_0$ and $V_0$, the pulsed enhancements described in this work are only available for certain regions of the ($\alpha_0$,$V_0$) parameter space. Although it has already been mentioned that this effect is limited to the low damping regime, there is the further simultaneous requirement that $V_0$ is not so high that leakage is not a concern. This $V_0$ decreases with increasing $\alpha_0$, as shown in Fig. \ref{fig:leak}. 

\begin{figure}[h]
\centering     
\includegraphics[width=\linewidth]{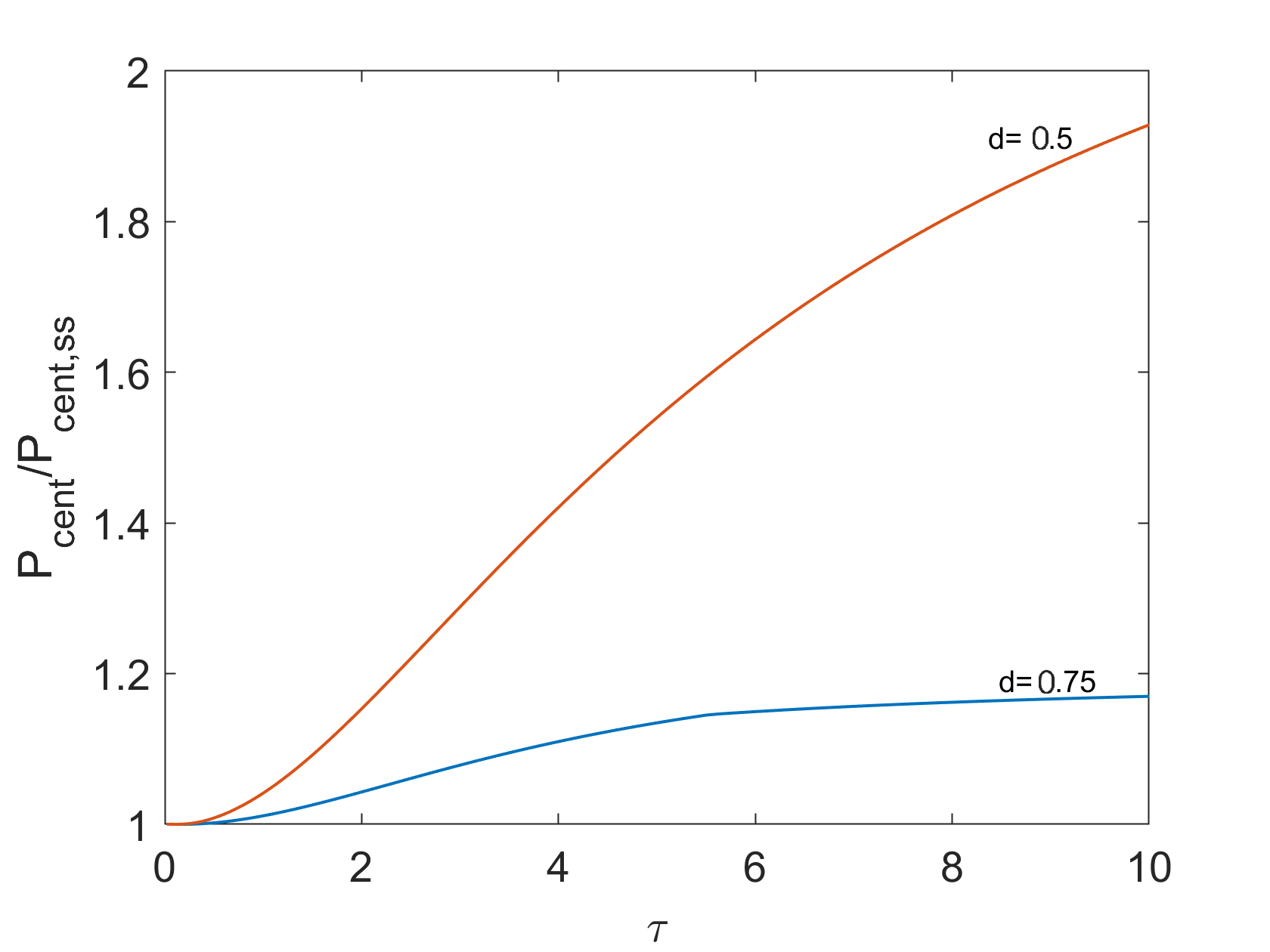}
\caption{\label{fig:nooptimum}Relative $P_{cent}$ improvements for non optimizable cases.}
\end{figure}

There are also similarly understood accessibility requirements on the duty cycle $d$, in that a high enough instantaneous power must be used to access a region of $\alpha_0,V_0$ space without leakage. Although some improvement will be seen, as any additional power still goes towards mitigating leakage, the full potential of the effect will not be realized and there will be no optimum pulsing frequency. For the narrow band of instantaneous powers that are high enough to just barely overcome leakage but not encounter shadowing, there is a strong performance enhancement from pulsing, but no optimum pulsing frequency. For these cases, performance monotonically improves with pulse times as the instantaneous parameters correspond to a highly favorable steady state, but then additional physics must be taken into account such as heat loss to ions. Examples of such non-optimizable cases are shown for the $\alpha_0=.2, V_0=4 $ case in Fig. \ref{fig:nooptimum}.
The accessibility requirements on $\alpha_0, V_0$ mentioned here are feasible for present devices, for which $\alpha_0\sim0.1-3, V_0 \sim 10$. \cite{Rodriguez_2019}

\section{Discussion}
It has been demonstrated that pulsed rf 
stabilization schemes can achieve significantly improved heating and stabilization compared to steady-state schemes of equivalent average power,
in regimes where the deposition width is large relative to the island size. The improvement is optimized when the pulsing time is on the order of the heat diffusion time. This effect can be understood as using higher instantaneous powers to achieve sufficient heating for nonlinear deposition enhancement to overcome the power leakage that would be present in the equivalent steady state. As the heating is nonlinearly enhanced while cooling remains a slow linear process, significantly higher cycle averaged temperatures and stabilizing current can be achieved for the same steady state power. The centralized heating and current profiles despite peripheral rf entry are only made possible by the diffusive nature of the island--pulsing on this natural time scale allows for active exploitation of this property. 

The caveats for the accessibility of the pulsed enhancements suggest that the effect explored here would have particular utility for LHCD, where the deposition profiles tend to be broader than those for ECCD. Such broad deposition profiles are desirable for the operation of a tokamak in steady state. In that scenario, the deposition may be locally enhanced in any islands that appear via the current condensation effect, providing automatic stabilization without controlled aiming of ray trajectories. Pulsing on typical island diffusive time scales may then be employed to improve power efficiency. Additionally, it was shown that high enough powers are required to overcome leakage, but carry a risk of shadowing. Considering such parameter sensitivity and generally high uncertainty of such parameters, the results suggest that in practice it may be best keep $\tau_{on}$ safely under a diffusion time, rather than aiming for the theoretical optimum. It must also be noted that a highly simplified model has been used here, and as such provides insights 

The trends predicted here for how the optimum pulsing frequency should depend on plasma and island parameters invite experimental verification. As the modulation frequency is swept from above diffusive time scales (> a few kHz, well within the reach of present devices) to below (~10 Hz), the cycle averaged island temperature should increase from the steady state until a maximum and then decrease, eventually heating less effectively than the equivalent steady state only when the frequencies fall below the order of diffusive time scales. This suggests that it will be safer to overshoot the pulsing frequency, as it may be difficult in practice to determine the precise optimum pulsing frequency. Fortunately, this effect is not sensitive to establishing a precise resonance. The performance deteriorates significantly only on the slower half of frequency space. This highlights the robustness with respect to frequency of the pulsing strategies as described here. In contrast, present modulation strategies for stabilizing rotating islands rely on precise matching of island rotation rates and phasing. 

In present devices, typical diffusion times correspond to frequencies of $\sim$1kHz, coincidentally right around the natural island rotation rates to which pulse times are matched. The majority of modulation experiments report modest improvements, with stabilization made possible at lower average powers compared to continuous wave schemes \cite{Maraschek_2007, Volpe_2009, Westerhof_2007, Isayama_2009,Kasparek_2016}. It has also been found that modulation seems to provide more a benefit when deposition is broad \cite{Maraschek_2007,Zohm_2007}. These results have been interpreted thus far as consequences of having the deposition better coincide with the O-point, but it is possible that the effects described here may have contributed to such success. Existing experiments only allow for speculation as to how large of a role the nonlinear effect has been playing. To resolve this ambiguity, an ideal experiment for testing this diffusion based modulation method would be using resonant magnetic perturbations (RMPs) to ensure O-point deposition, while sweeping the modulation frequency and duty cycle for a fixed average power to allow direct comparison to the corresponding continuous wave scheme. The anticipated experimental signature would be clear--island temperature increasing with pulse times until an optimum is observed. 

While the full treatment of ion temperature is left for future work, the limiting cases presented in appendix B motivate the investigation of pulsing as a means of limiting parasitic heat loss to the ions. When electrons and ions fully share the input power, the effective $V_0$ is reduced by a factor of $(1+\kappa_i/\kappa_e)$. Considering that only the electron temperature contributes to the nonlinear effect for the waves considered here, this amounts to a considerable loss in stabilization capability for a given amount of input power. Additional inhibition mechanisms such as stiffness\cite{eduardo}, triggered by ion temperature gradients, further motivate this line of inquiry.

The potential for pulsed rf schemes in the high damping regime is also left for future work. The high damping regime is characterized by flat-topped temperature profiles and is prone to shadowing-- even at low powers, any additional power input only contributes to even more peripheral deposition, damaging stabilization efforts. The potential for exploiting diffusion in an arguably more direct way with a ``cooling-based" method, can be understood as follows. Given a flat-topped shadowed temperature profile, upon cessation of heating, temperature will be lost rapidly from the periphery but slowly in the center, creating a more favorable damping landscape for a ray entering after this cooling period. Such an effect is present but offset by sensitivity to shadowing in this 1-D slab model, but could become dominant with a more accurate treatment of island geometry. It is further anticipated that multiple rays (with different $\alpha_0$) may be synergistic and especially successful in this high damping regime, a possibility also left for future work.

\section{Summary}
Pulsing rf power on diffusive time scales has the potential to achieve significantly improved heating and stabilization of magnetic islands for the same cycle-averaged power. This has been explicitly demonstrated using a simple 1-D coupled wave damping- island temperature diffusion model in the low damping regime, marked by poor linear absorption. Such pulsed schemes exploit nonlinear heating and the diffusive nature of the island temperature to overcome power leakage while avoiding shadowing. The optimum pulse time is anticipated to be on the order of the diffusion time, and increase with increasing duty cycle or decreasing damping strength $\alpha_0$ or cycle-averaged power $V_{0,eff}$. These predicted trends lend themselves naturally to experimental verification. Interestingly, diffusion times happen to be on the order of modulation times in experiments that pulse to rf to match island rotation. This opens the possibility that the effects described here could have contributed to the performance of those modulated schemes, further motivating experiments to untangle the effects of island-phase matching and the nonlinear heating.
\begin{acknowledgements}
Thanks to Eduardo Rodriguez for helpful discussions.

This work was supported by Nos. U.S. DOE DE-AC02-
09CH11466 and DE-SC0016072.
\end{acknowledgements}
\appendix
\section{Figures of merit for optimizing pulsing strategies}

\begin{figure}[h]
\centering     
\subfigure{\label{fig:a}\includegraphics[width=\linewidth]{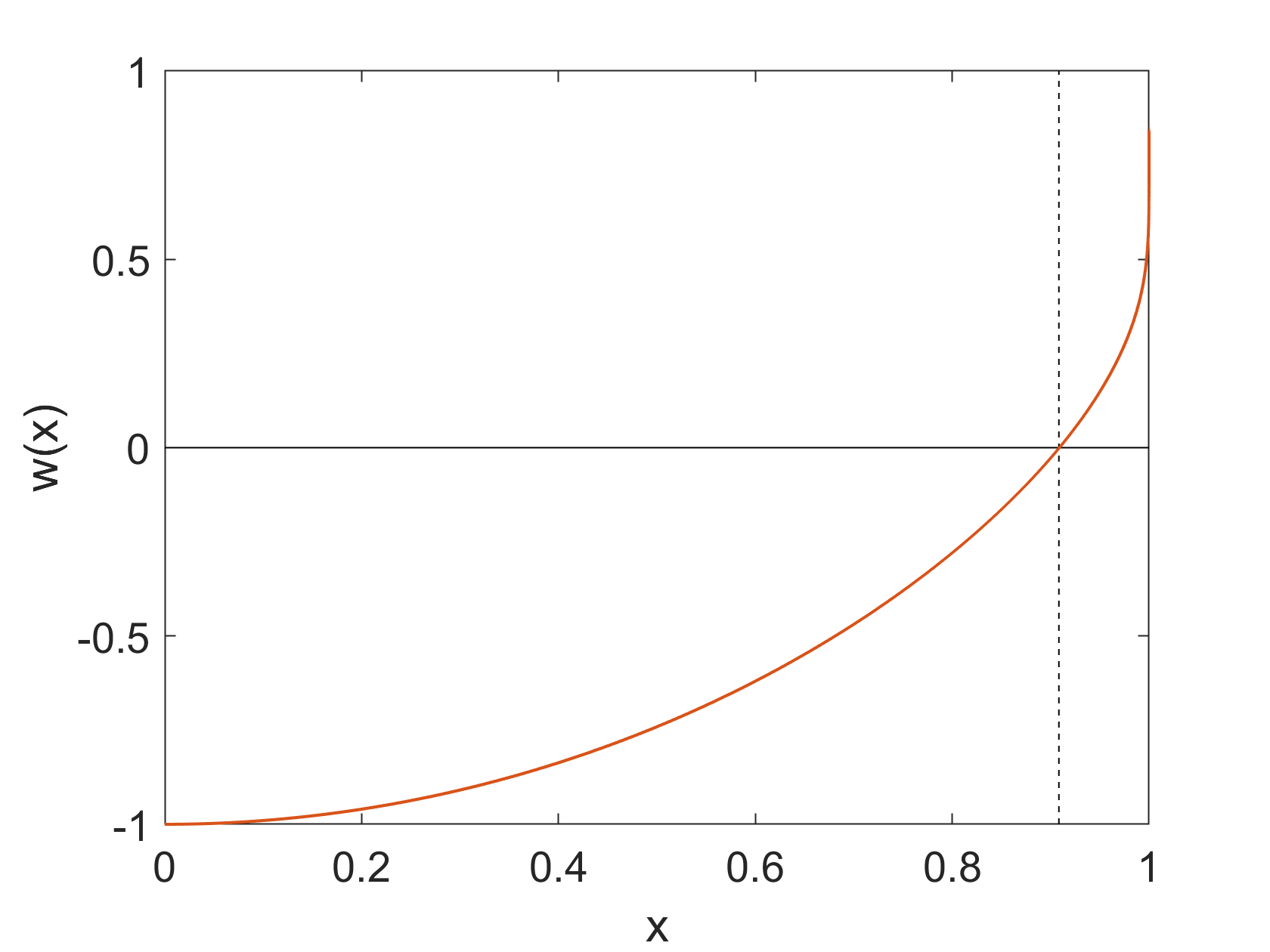}}
\subfigure  {\label{fig:b}\includegraphics[width=\linewidth]{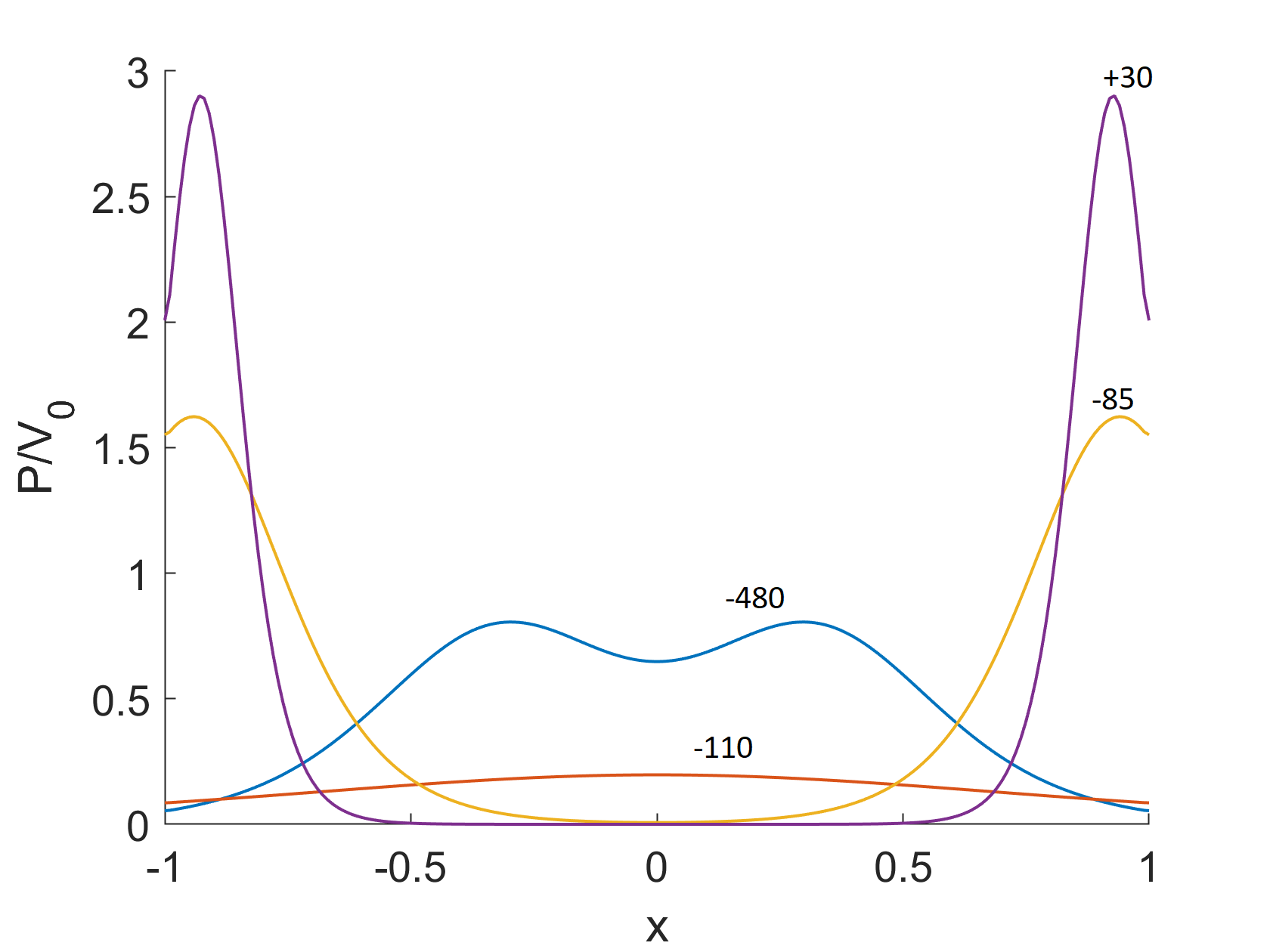}}
\caption{Various power deposition profiles and their corresponding stability ratings (b), evaluated using the weight function (a). The power profiles are normalized to the input power, but the stability values $\sigma$ are not. }
\end{figure}
While traditional stability calculations are inherently 2-D in the poloidal plane, a 1-D analog can be defined: $\sigma:=\int_0^1 w(x) P(x) dx$ where w(x) is a weight function obtained using a generic island geometry. $\sigma$ is proportional to the current drive contribution to the island growth rate, with negative values indicating stabilizing scenarios and vice versa. The weight function $w(x)$ is shown in Figure 2 along with some example power deposition profiles and their sigma values. The derivation for the specific form of w(x) can be found in Ref. \onlinecite{eduardo}.

Evidently, power driven in the outer 10\% of the island is destabilizing and as such, $\sigma$ is too insensitive of a metric for the purpose of studying how pulse parameters affect deposition. Additionally, as $\tau$ is increased $\sigma$ very slowly reaches a maximum, then sharply drops--undershooting is always far safer. This motivates the introduction of the more sensitive $P_{cent}=\int _{-.5}^{.5} P dx/ V_0$, which gives the fraction of available power deposited in the center half of the island. $P_{cent}$ retains the spirit of favoring central deposition while providing a sharper objective function. 
\section{Heat loss to ions}
The electron temperature equation (1) as written, carries the implicit assumption that the electron diffusion time $t_{D,e}$ is much faster than the electron-ion equilibration time $t_{eq}$. This can be seen by starting from the 1-D linearized two fluid equations (with $Z=1,\quad n_e=n_i=n$ for simplicity):
\begin{equation}
     \frac{3}{2} n \partial_t T_e - \kappa_\perp^e \partial_x^2 T_e=P+n \nu (T_i-T_e)
\end{equation}
\begin{equation}
     \frac{3}{2} n \partial_t T_i - \kappa_\perp^i \partial_x^2 T_e=n \nu (T_e-T_i)
\end{equation}

Now employing the same scalings used in section II, the equations become:
\begin{equation}
    \dot u_e - u_e''= p +c(u_i-u_e)
\end{equation}
\begin{equation}
    \dot u_i - \gamma u_i''= c(u_e-u_i)
\end{equation}
where $p= -(V'(x)+V'(-x))/2 $ as before, $c := 2 t_{D,e}/3 t_{eq}$ with $t_{D,e}=3 W_i/8 \chi_\perp^e$ , $t_{eq}=1/\nu$

Obviously, as $c \rightarrow0$, our equations reduce to the single fluid electron heating, cold ion model used in the rest of this work. However, rearranging the ion equation (B4) to the more suggestive form:
\begin{equation}
    u_i=u_e-\frac{1}{c}(\dot u_i - \gamma u_i'')
\end{equation}
shows that as $c \rightarrow \infty$, we can take $u_i \approx u_e=u$ with only small corrections. 

Adding the ion and electron equations gives
\begin{equation}
    \dot u - \frac{1+\gamma}{2} u''= p/2
\end{equation}
so the original problem is recovered with a new conductivity that is the average of the electron and ion conductivities, and a halved source term. The exact form of equation 6 is then obtained by substituting new scalings ($t_{scl} \rightarrow t_{scl} 2/(1+\gamma) ,  \quad V_{scl} \rightarrow V_{scl} (1+\gamma)$. We see that in this limit, the power is effective reduced by the factor $1/(1+\gamma)$. 
A mixed Bohm/gyro-Bohm model for heat transport gives $\gamma \approx 2$.\cite{Erba_1998} 

 For electron diffusivities in the range $\chi_e \sim 0.1 - 1\text{ m}^2\text{/s } $ \cite{Westerhof_2007} and typical tokamak parameters on the order of $n\sim10^{20} \text{m}^{-3}$, $\quad T\sim1 \text{ keV}, \quad W_i/2\sim 1 \text{cm}, \quad B \sim 1 \text{ T}$, the electron-ion equilibration time is on the order of $t_{eq} \sim 10^-2 \text{ s}$ and electron diffusion time $t_{D,e}\sim 10^{-4} - 10^{-3} \text{ s}$. 
 These estimates give $c \sim 10^{-2}-10^{-1}$, so the electron temperature alone is considered in this work. Adopting the $\chi_\perp^e$ scalings used in Ref. \onlinecite{Erba_1998}, we get that $c\propto n B W_i^2/T^{2.5}$, so the relative importance of heat loss to ions is expected to increase for large islands in strong magnetic fields.
 
\nocite{*}

\bibliography{aipsamp}

\providecommand{\noopsort}[1]{}\providecommand{\singleletter}[1]{#1}%
\begin{thebibliography}{62}%
\makeatletter
\providecommand \@ifxundefined [1]{%
 \@ifx{#1\undefined}
}%
\providecommand \@ifnum [1]{%
 \ifnum #1\expandafter \@firstoftwo
 \else \expandafter \@secondoftwo
 \fi
}%
\providecommand \@ifx [1]{%
 \ifx #1\expandafter \@firstoftwo
 \else \expandafter \@secondoftwo
 \fi
}%
\providecommand \natexlab [1]{#1}%
\providecommand \enquote  [1]{``#1''}%
\providecommand \bibnamefont  [1]{#1}%
\providecommand \bibfnamefont [1]{#1}%
\providecommand \citenamefont [1]{#1}%
\providecommand \href@noop [0]{\@secondoftwo}%
\providecommand \href [0]{\begingroup \@sanitize@url \@href}%
\providecommand \@href[1]{\@@startlink{#1}\@@href}%
\providecommand \@@href[1]{\endgroup#1\@@endlink}%
\providecommand \@sanitize@url [0]{\catcode `\\12\catcode `\$12\catcode
  `\&12\catcode `\#12\catcode `\^12\catcode `\_12\catcode `\%12\relax}%
\providecommand \@@startlink[1]{}%
\providecommand \@@endlink[0]{}%
\providecommand \url  [0]{\begingroup\@sanitize@url \@url }%
\providecommand \@url [1]{\endgroup\@href {#1}{\urlprefix }}%
\providecommand \urlprefix  [0]{URL }%
\providecommand \Eprint [0]{\href }%
\providecommand \doibase [0]{http://dx.doi.org/}%
\providecommand \selectlanguage [0]{\@gobble}%
\providecommand \bibinfo  [0]{\@secondoftwo}%
\providecommand \bibfield  [0]{\@secondoftwo}%
\providecommand \translation [1]{[#1]}%
\providecommand \BibitemOpen [0]{}%
\providecommand \bibitemStop [0]{}%
\providecommand \bibitemNoStop [0]{.\EOS\space}%
\providecommand \EOS [0]{\spacefactor3000\relax}%
\providecommand \BibitemShut  [1]{\csname bibitem#1\endcsname}%
\let\auto@bib@innerbib\@empty
\bibitem [{\citenamefont {de~Vries}\ \emph {et~al.}(2011)\citenamefont
  {de~Vries}, \citenamefont {Johnson}, \citenamefont {Alper}, \citenamefont
  {Buratti}, \citenamefont {Hender}, \citenamefont {Koslowski},\ and\
  \citenamefont {and}}]{de_Vries_2011}%
  \BibitemOpen
  \bibfield  {author} {\bibinfo {author} {\bibfnamefont {P.}~\bibnamefont
  {de~Vries}}, \bibinfo {author} {\bibfnamefont {M.}~\bibnamefont {Johnson}},
  \bibinfo {author} {\bibfnamefont {B.}~\bibnamefont {Alper}}, \bibinfo
  {author} {\bibfnamefont {P.}~\bibnamefont {Buratti}}, \bibinfo {author}
  {\bibfnamefont {T.}~\bibnamefont {Hender}}, \bibinfo {author} {\bibfnamefont
  {H.}~\bibnamefont {Koslowski}}, \ and\ \bibinfo {author} {\bibfnamefont
  {V.~R.}\ \bibnamefont {and}},\ }\bibfield  {title} {\enquote {\bibinfo
  {title} {Survey of disruption causes at {JET}},}\ }\href {\doibase
  10.1088/0029-5515/51/5/053018} {\bibfield  {journal} {\bibinfo  {journal}
  {Nuclear Fusion}\ }\textbf {\bibinfo {volume} {51}},\ \bibinfo {pages}
  {053018} (\bibinfo {year} {2011})}\BibitemShut {NoStop}%
\bibitem [{\citenamefont {de~Vries}\ \emph {et~al.}(2014)\citenamefont
  {de~Vries}, \citenamefont {Baruzzo}, \citenamefont {Hogeweij}, \citenamefont
  {Jachmich}, \citenamefont {Joffrin}, \citenamefont {Lomas}, \citenamefont
  {Matthews}, \citenamefont {Murari}, \citenamefont {Nunes}, \citenamefont
  {Pütterich}, \citenamefont {Reux},\ and\ \citenamefont
  {Vega}}]{de_Vries_2014}%
  \BibitemOpen
  \bibfield  {author} {\bibinfo {author} {\bibfnamefont {P.~C.}\ \bibnamefont
  {de~Vries}}, \bibinfo {author} {\bibfnamefont {M.}~\bibnamefont {Baruzzo}},
  \bibinfo {author} {\bibfnamefont {G.~M.~D.}\ \bibnamefont {Hogeweij}},
  \bibinfo {author} {\bibfnamefont {S.}~\bibnamefont {Jachmich}}, \bibinfo
  {author} {\bibfnamefont {E.}~\bibnamefont {Joffrin}}, \bibinfo {author}
  {\bibfnamefont {P.~J.}\ \bibnamefont {Lomas}}, \bibinfo {author}
  {\bibfnamefont {G.~F.}\ \bibnamefont {Matthews}}, \bibinfo {author}
  {\bibfnamefont {A.}~\bibnamefont {Murari}}, \bibinfo {author} {\bibfnamefont
  {I.}~\bibnamefont {Nunes}}, \bibinfo {author} {\bibfnamefont
  {T.}~\bibnamefont {Pütterich}}, \bibinfo {author} {\bibfnamefont
  {C.}~\bibnamefont {Reux}}, \ and\ \bibinfo {author} {\bibfnamefont
  {J.}~\bibnamefont {Vega}},\ }\bibfield  {title} {\enquote {\bibinfo {title}
  {The influence of an iter-like wall on disruptions at jet},}\ }\href
  {\doibase 10.1063/1.4872017} {\bibfield  {journal} {\bibinfo  {journal}
  {Physics of Plasmas}\ }\textbf {\bibinfo {volume} {21}},\ \bibinfo {pages}
  {056101} (\bibinfo {year} {2014})}\BibitemShut {NoStop}%
\bibitem [{\citenamefont {Haye}\ \emph {et~al.}(2006)\citenamefont {Haye},
  \citenamefont {Prater}, \citenamefont {Buttery}, \citenamefont {Hayashi},
  \citenamefont {Isayama}, \citenamefont {Maraschek}, \citenamefont {Urso},\
  and\ \citenamefont {Zohm}}]{La_Haye_2006a}%
  \BibitemOpen
  \bibfield  {author} {\bibinfo {author} {\bibfnamefont {R.~L.}\ \bibnamefont
  {Haye}}, \bibinfo {author} {\bibfnamefont {R.}~\bibnamefont {Prater}},
  \bibinfo {author} {\bibfnamefont {R.}~\bibnamefont {Buttery}}, \bibinfo
  {author} {\bibfnamefont {N.}~\bibnamefont {Hayashi}}, \bibinfo {author}
  {\bibfnamefont {A.}~\bibnamefont {Isayama}}, \bibinfo {author} {\bibfnamefont
  {M.}~\bibnamefont {Maraschek}}, \bibinfo {author} {\bibfnamefont
  {L.}~\bibnamefont {Urso}}, \ and\ \bibinfo {author} {\bibfnamefont
  {H.}~\bibnamefont {Zohm}},\ }\bibfield  {title} {\enquote {\bibinfo {title}
  {Cross{\textendash}machine benchmarking for {ITER} of neoclassical tearing
  mode stabilization by electron cyclotron current drive},}\ }\href {\doibase
  10.1088/0029-5515/46/4/006} {\bibfield  {journal} {\bibinfo  {journal}
  {Nuclear Fusion}\ }\textbf {\bibinfo {volume} {46}},\ \bibinfo {pages}
  {451--461} (\bibinfo {year} {2006})}\BibitemShut {NoStop}%
\bibitem [{\citenamefont {Haye}, \citenamefont {Isayama},\ and\ \citenamefont
  {Maraschek}(2009)}]{La_Haye_2009}%
  \BibitemOpen
  \bibfield  {author} {\bibinfo {author} {\bibfnamefont {R.~L.}\ \bibnamefont
  {Haye}}, \bibinfo {author} {\bibfnamefont {A.}~\bibnamefont {Isayama}}, \
  and\ \bibinfo {author} {\bibfnamefont {M.}~\bibnamefont {Maraschek}},\
  }\bibfield  {title} {\enquote {\bibinfo {title} {Prospects for stabilization
  of neoclassical tearing modes by electron cyclotron current drive in
  {ITER}},}\ }\href {\doibase 10.1088/0029-5515/49/4/045005} {\bibfield
  {journal} {\bibinfo  {journal} {Nuclear Fusion}\ }\textbf {\bibinfo {volume}
  {49}},\ \bibinfo {pages} {045005} (\bibinfo {year} {2009})}\BibitemShut
  {NoStop}%
\bibitem [{\citenamefont {Reiman}(1983)}]{Reiman_1983}%
  \BibitemOpen
  \bibfield  {author} {\bibinfo {author} {\bibfnamefont {A.~H.}\ \bibnamefont
  {Reiman}},\ }\bibfield  {title} {\enquote {\bibinfo {title} {Suppression of
  magnetic islands by rf driven currents},}\ }\href {\doibase 10.1063/1.864258}
  {\bibfield  {journal} {\bibinfo  {journal} {The Physics of Fluids}\ }\textbf
  {\bibinfo {volume} {26}},\ \bibinfo {pages} {1338--1340} (\bibinfo {year}
  {1983})}\BibitemShut {NoStop}%
\bibitem [{\citenamefont {Zohm}(1997)}]{Zohm_1997}%
  \BibitemOpen
  \bibfield  {author} {\bibinfo {author} {\bibfnamefont {H.}~\bibnamefont
  {Zohm}},\ }\bibfield  {title} {\enquote {\bibinfo {title} {Stabilization of
  neoclassical tearing modes by electron cyclotron current drive},}\ }\href
  {\doibase 10.1063/1.872487} {\bibfield  {journal} {\bibinfo  {journal}
  {Physics of Plasmas}\ }\textbf {\bibinfo {volume} {4}},\ \bibinfo {pages}
  {3433--3435} (\bibinfo {year} {1997})}\BibitemShut {NoStop}%
\bibitem [{\citenamefont {Yu}\ and\ \citenamefont {Günter}(1998)}]{Yu_1998}%
  \BibitemOpen
  \bibfield  {author} {\bibinfo {author} {\bibfnamefont {Q.}~\bibnamefont
  {Yu}}\ and\ \bibinfo {author} {\bibfnamefont {S.}~\bibnamefont {Günter}},\
  }\bibfield  {title} {\enquote {\bibinfo {title} {On the stabilization of
  neoclassical tearing modes by phased electron cyclotron waves},}\ }\href
  {\doibase 10.1088/0741-3335/40/11/011} {\bibfield  {journal} {\bibinfo
  {journal} {Plasma Physics and Controlled Fusion}\ }\textbf {\bibinfo {volume}
  {40}},\ \bibinfo {pages} {1977--1987} (\bibinfo {year} {1998})}\BibitemShut
  {NoStop}%
\bibitem [{\citenamefont {Harvey}\ and\ \citenamefont
  {Perkins}(2001)}]{Harvey_2001}%
  \BibitemOpen
  \bibfield  {author} {\bibinfo {author} {\bibfnamefont {R.}~\bibnamefont
  {Harvey}}\ and\ \bibinfo {author} {\bibfnamefont {F.}~\bibnamefont
  {Perkins}},\ }\bibfield  {title} {\enquote {\bibinfo {title} {Comparison of
  optimized {ECCD} for different launch locations in a next step tokamak
  reactor plasma},}\ }\href {\doibase 10.1088/0029-5515/41/12/312} {\bibfield
  {journal} {\bibinfo  {journal} {Nuclear Fusion}\ }\textbf {\bibinfo {volume}
  {41}},\ \bibinfo {pages} {1847--1856} (\bibinfo {year} {2001})}\BibitemShut
  {NoStop}%
\bibitem [{\citenamefont {Bernabei}\ \emph {et~al.}(1998)\citenamefont
  {Bernabei}, \citenamefont {Cardinali}, \citenamefont {Giruzzi},\ and\
  \citenamefont {Zabi{\'{e}}go}}]{Bernabei_1998}%
  \BibitemOpen
  \bibfield  {author} {\bibinfo {author} {\bibfnamefont {S.}~\bibnamefont
  {Bernabei}}, \bibinfo {author} {\bibfnamefont {A.}~\bibnamefont {Cardinali}},
  \bibinfo {author} {\bibfnamefont {G.}~\bibnamefont {Giruzzi}}, \ and\
  \bibinfo {author} {\bibfnamefont {M.}~\bibnamefont {Zabi{\'{e}}go}},\
  }\bibfield  {title} {\enquote {\bibinfo {title} {Tearing mode stabilization
  in tokamaks with lower hybrid waves},}\ }\href {\doibase
  10.1088/0029-5515/38/1/307} {\bibfield  {journal} {\bibinfo  {journal}
  {Nuclear Fusion}\ }\textbf {\bibinfo {volume} {38}},\ \bibinfo {pages}
  {87--92} (\bibinfo {year} {1998})}\BibitemShut {NoStop}%
\bibitem [{\citenamefont {Kamendje}\ \emph {et~al.}(2005)\citenamefont
  {Kamendje}, \citenamefont {Kasilov}, \citenamefont {Kernbichler},
  \citenamefont {Pavlenko}, \citenamefont {Poli},\ and\ \citenamefont
  {Heyn}}]{Kamendje_2005}%
  \BibitemOpen
  \bibfield  {author} {\bibinfo {author} {\bibfnamefont {R.}~\bibnamefont
  {Kamendje}}, \bibinfo {author} {\bibfnamefont {S.~V.}\ \bibnamefont
  {Kasilov}}, \bibinfo {author} {\bibfnamefont {W.}~\bibnamefont
  {Kernbichler}}, \bibinfo {author} {\bibfnamefont {I.~V.}\ \bibnamefont
  {Pavlenko}}, \bibinfo {author} {\bibfnamefont {E.}~\bibnamefont {Poli}}, \
  and\ \bibinfo {author} {\bibfnamefont {M.~F.}\ \bibnamefont {Heyn}},\
  }\bibfield  {title} {\enquote {\bibinfo {title} {Modeling of nonlinear
  electron cyclotron resonance heating and current drive in a tokamak},}\
  }\href {\doibase 10.1063/1.1823415} {\bibfield  {journal} {\bibinfo
  {journal} {Physics of Plasmas}\ }\textbf {\bibinfo {volume} {12}},\ \bibinfo
  {pages} {012502} (\bibinfo {year} {2005})}\BibitemShut {NoStop}%
\bibitem [{\citenamefont {La~Haye}(2006)}]{La_Haye_2006b}%
  \BibitemOpen
  \bibfield  {author} {\bibinfo {author} {\bibfnamefont {R.~J.}\ \bibnamefont
  {La~Haye}},\ }\bibfield  {title} {\enquote {\bibinfo {title} {Neoclassical
  tearing modes and their control},}\ }\href {\doibase 10.1063/1.2180747}
  {\bibfield  {journal} {\bibinfo  {journal} {Physics of Plasmas}\ }\textbf
  {\bibinfo {volume} {13}},\ \bibinfo {pages} {055501} (\bibinfo {year}
  {2006})}\BibitemShut {NoStop}%
\bibitem [{\citenamefont {Haye}\ \emph {et~al.}(2008)\citenamefont {Haye},
  \citenamefont {Ferron}, \citenamefont {Humphreys}, \citenamefont {Luce},
  \citenamefont {Petty}, \citenamefont {Prater}, \citenamefont {Strait},\ and\
  \citenamefont {Welander}}]{La_Haye_2008}%
  \BibitemOpen
  \bibfield  {author} {\bibinfo {author} {\bibfnamefont {R.~L.}\ \bibnamefont
  {Haye}}, \bibinfo {author} {\bibfnamefont {J.}~\bibnamefont {Ferron}},
  \bibinfo {author} {\bibfnamefont {D.}~\bibnamefont {Humphreys}}, \bibinfo
  {author} {\bibfnamefont {T.}~\bibnamefont {Luce}}, \bibinfo {author}
  {\bibfnamefont {C.}~\bibnamefont {Petty}}, \bibinfo {author} {\bibfnamefont
  {R.}~\bibnamefont {Prater}}, \bibinfo {author} {\bibfnamefont
  {E.}~\bibnamefont {Strait}}, \ and\ \bibinfo {author} {\bibfnamefont
  {A.}~\bibnamefont {Welander}},\ }\bibfield  {title} {\enquote {\bibinfo
  {title} {Requirements for alignment of electron cyclotron current drive for
  neoclassical tearing mode stabilization in {ITER}},}\ }\href {\doibase
  10.1088/0029-5515/48/5/054004} {\bibfield  {journal} {\bibinfo  {journal}
  {Nuclear Fusion}\ }\textbf {\bibinfo {volume} {48}},\ \bibinfo {pages}
  {054004} (\bibinfo {year} {2008})}\BibitemShut {NoStop}%
\bibitem [{\citenamefont {Sauter}\ \emph {et~al.}(2010)\citenamefont {Sauter},
  \citenamefont {Henderson}, \citenamefont {Ramponi}, \citenamefont {Zohm},\
  and\ \citenamefont {Zucca}}]{Sauter_2010}%
  \BibitemOpen
  \bibfield  {author} {\bibinfo {author} {\bibfnamefont {O.}~\bibnamefont
  {Sauter}}, \bibinfo {author} {\bibfnamefont {M.~A.}\ \bibnamefont
  {Henderson}}, \bibinfo {author} {\bibfnamefont {G.}~\bibnamefont {Ramponi}},
  \bibinfo {author} {\bibfnamefont {H.}~\bibnamefont {Zohm}}, \ and\ \bibinfo
  {author} {\bibfnamefont {C.}~\bibnamefont {Zucca}},\ }\bibfield  {title}
  {\enquote {\bibinfo {title} {On the requirements to control neoclassical
  tearing modes in burning plasmas},}\ }\href {\doibase
  10.1088/0741-3335/52/2/025002} {\bibfield  {journal} {\bibinfo  {journal}
  {Plasma Physics and Controlled Fusion}\ }\textbf {\bibinfo {volume} {52}},\
  \bibinfo {pages} {025002} (\bibinfo {year} {2010})}\BibitemShut {NoStop}%
\bibitem [{\citenamefont {Smolyakov}\ \emph {et~al.}(2013)\citenamefont
  {Smolyakov}, \citenamefont {Poye}, \citenamefont {Agullo}, \citenamefont
  {Benkadda},\ and\ \citenamefont {Garbet}}]{Smolyakov_2013}%
  \BibitemOpen
  \bibfield  {author} {\bibinfo {author} {\bibfnamefont {A.~I.}\ \bibnamefont
  {Smolyakov}}, \bibinfo {author} {\bibfnamefont {A.}~\bibnamefont {Poye}},
  \bibinfo {author} {\bibfnamefont {O.}~\bibnamefont {Agullo}}, \bibinfo
  {author} {\bibfnamefont {S.}~\bibnamefont {Benkadda}}, \ and\ \bibinfo
  {author} {\bibfnamefont {X.}~\bibnamefont {Garbet}},\ }\bibfield  {title}
  {\enquote {\bibinfo {title} {Higher order and asymmetry effects on saturation
  of magnetic islands},}\ }\href {\doibase 10.1063/1.4811383} {\bibfield
  {journal} {\bibinfo  {journal} {Physics of Plasmas}\ }\textbf {\bibinfo
  {volume} {20}},\ \bibinfo {pages} {062506} (\bibinfo {year}
  {2013})}\BibitemShut {NoStop}%
\bibitem [{\citenamefont {Ayten}\ and\ \citenamefont {and}(2014)}]{Ayten_2014}%
  \BibitemOpen
  \bibfield  {author} {\bibinfo {author} {\bibfnamefont {B.}~\bibnamefont
  {Ayten}}\ and\ \bibinfo {author} {\bibfnamefont {E.~W.}\ \bibnamefont
  {and}},\ }\bibfield  {title} {\enquote {\bibinfo {title} {Non-linear effects
  in electron cyclotron current drive applied for the stabilization of
  neoclassical tearing modes},}\ }\href {\doibase
  10.1088/0029-5515/54/7/073001} {\bibfield  {journal} {\bibinfo  {journal}
  {Nuclear Fusion}\ }\textbf {\bibinfo {volume} {54}},\ \bibinfo {pages}
  {073001} (\bibinfo {year} {2014})}\BibitemShut {NoStop}%
\bibitem [{\citenamefont {Borgogno}\ \emph {et~al.}(2014)\citenamefont
  {Borgogno}, \citenamefont {Comisso}, \citenamefont {Grasso},\ and\
  \citenamefont {Lazzaro}}]{Borgogno_2014}%
  \BibitemOpen
  \bibfield  {author} {\bibinfo {author} {\bibfnamefont {D.}~\bibnamefont
  {Borgogno}}, \bibinfo {author} {\bibfnamefont {L.}~\bibnamefont {Comisso}},
  \bibinfo {author} {\bibfnamefont {D.}~\bibnamefont {Grasso}}, \ and\ \bibinfo
  {author} {\bibfnamefont {E.}~\bibnamefont {Lazzaro}},\ }\bibfield  {title}
  {\enquote {\bibinfo {title} {Nonlinear response of magnetic islands to
  localized electron cyclotron current injection},}\ }\href {\doibase
  10.1063/1.4885635} {\bibfield  {journal} {\bibinfo  {journal} {Physics of
  Plasmas}\ }\textbf {\bibinfo {volume} {21}},\ \bibinfo {pages} {060704}
  (\bibinfo {year} {2014})}\BibitemShut {NoStop}%
\bibitem [{\citenamefont {F{\'{e}}vrier}\ \emph {et~al.}(2016)\citenamefont
  {F{\'{e}}vrier}, \citenamefont {Maget}, \citenamefont {Lütjens},
  \citenamefont {Luciani}, \citenamefont {Decker}, \citenamefont {Giruzzi},
  \citenamefont {Reich}, \citenamefont {Beyer}, \citenamefont {Lazzaro},\ and\
  \citenamefont {and}}]{F_vrier_2016}%
  \BibitemOpen
  \bibfield  {author} {\bibinfo {author} {\bibfnamefont {O.}~\bibnamefont
  {F{\'{e}}vrier}}, \bibinfo {author} {\bibfnamefont {P.}~\bibnamefont
  {Maget}}, \bibinfo {author} {\bibfnamefont {H.}~\bibnamefont {Lütjens}},
  \bibinfo {author} {\bibfnamefont {J.~F.}\ \bibnamefont {Luciani}}, \bibinfo
  {author} {\bibfnamefont {J.}~\bibnamefont {Decker}}, \bibinfo {author}
  {\bibfnamefont {G.}~\bibnamefont {Giruzzi}}, \bibinfo {author} {\bibfnamefont
  {M.}~\bibnamefont {Reich}}, \bibinfo {author} {\bibfnamefont
  {P.}~\bibnamefont {Beyer}}, \bibinfo {author} {\bibfnamefont
  {E.}~\bibnamefont {Lazzaro}}, \ and\ \bibinfo {author} {\bibfnamefont
  {S.~N.}\ \bibnamefont {and}},\ }\bibfield  {title} {\enquote {\bibinfo
  {title} {First principles fluid modelling of magnetic island stabilization by
  electron cyclotron current drive ({ECCD})},}\ }\href {\doibase
  10.1088/0741-3335/58/4/045015} {\bibfield  {journal} {\bibinfo  {journal}
  {Plasma Physics and Controlled Fusion}\ }\textbf {\bibinfo {volume} {58}},\
  \bibinfo {pages} {045015} (\bibinfo {year} {2016})}\BibitemShut {NoStop}%
\bibitem [{\citenamefont {Li}\ \emph {et~al.}(2017)\citenamefont {Li},
  \citenamefont {Xiao}, \citenamefont {Lin},\ and\ \citenamefont
  {Wang}}]{Li_2017}%
  \BibitemOpen
  \bibfield  {author} {\bibinfo {author} {\bibfnamefont {J.~C.}\ \bibnamefont
  {Li}}, \bibinfo {author} {\bibfnamefont {C.~J.}\ \bibnamefont {Xiao}},
  \bibinfo {author} {\bibfnamefont {Z.~H.}\ \bibnamefont {Lin}}, \ and\
  \bibinfo {author} {\bibfnamefont {K.~J.}\ \bibnamefont {Wang}},\ }\bibfield
  {title} {\enquote {\bibinfo {title} {Effects of electron cyclotron current
  drive on magnetic islands in tokamak plasmas},}\ }\href {\doibase
  10.1063/1.4996021} {\bibfield  {journal} {\bibinfo  {journal} {Physics of
  Plasmas}\ }\textbf {\bibinfo {volume} {24}},\ \bibinfo {pages} {082508}
  (\bibinfo {year} {2017})}\BibitemShut {NoStop}%
\bibitem [{\citenamefont {Grasso}\ \emph {et~al.}(2018)\citenamefont {Grasso},
  \citenamefont {Borgogno}, \citenamefont {Comisso},\ and\ \citenamefont
  {Lazzaro}}]{grasso_borgogno_comisso_lazzaro_2018}%
  \BibitemOpen
  \bibfield  {author} {\bibinfo {author} {\bibfnamefont {D.}~\bibnamefont
  {Grasso}}, \bibinfo {author} {\bibfnamefont {D.}~\bibnamefont {Borgogno}},
  \bibinfo {author} {\bibfnamefont {L.}~\bibnamefont {Comisso}}, \ and\
  \bibinfo {author} {\bibfnamefont {E.}~\bibnamefont {Lazzaro}},\ }\bibfield
  {title} {\enquote {\bibinfo {title} {Magnetic island suppression by
  electron cyclotron current drive as converse of a forced reconnection
  problem},}\ }\href {\doibase 10.1017/S0022377818000569} {\bibfield  {journal}
  {\bibinfo  {journal} {Journal of Plasma Physics}\ }\textbf {\bibinfo {volume}
  {84}},\ \bibinfo {pages} {745840302} (\bibinfo {year} {2018})}\BibitemShut
  {NoStop}%
\bibitem [{\citenamefont {Grasso}\ \emph {et~al.}(2016)\citenamefont {Grasso},
  \citenamefont {Lazzaro}, \citenamefont {Borgogno},\ and\ \citenamefont
  {Comisso}}]{Grasso_2016}%
  \BibitemOpen
  \bibfield  {author} {\bibinfo {author} {\bibfnamefont {D.}~\bibnamefont
  {Grasso}}, \bibinfo {author} {\bibfnamefont {E.}~\bibnamefont {Lazzaro}},
  \bibinfo {author} {\bibfnamefont {D.}~\bibnamefont {Borgogno}}, \ and\
  \bibinfo {author} {\bibfnamefont {L.}~\bibnamefont {Comisso}},\ }\bibfield
  {title} {\enquote {\bibinfo {title} {Open problems of magnetic island control
  by electron cyclotron current drive},}\ }\href {\doibase
  10.1017/S0022377816001045} {\bibfield  {journal} {\bibinfo  {journal}
  {Journal of Plasma Physics}\ }\textbf {\bibinfo {volume} {82}},\ \bibinfo
  {pages} {595820603} (\bibinfo {year} {2016})}\BibitemShut {NoStop}%
\bibitem [{\citenamefont {Zohm}\ \emph {et~al.}(1999)\citenamefont {Zohm},
  \citenamefont {Gantenbein}, \citenamefont {Giruzzi}, \citenamefont {Günter},
  \citenamefont {Leuterer}, \citenamefont {Maraschek}, \citenamefont {Meskat},
  \citenamefont {Peeters}, \citenamefont {Suttrop}, \citenamefont {Wagner},
  \citenamefont {Zabi{\'{e}}go}, \citenamefont {Team},\ and\ \citenamefont
  {Group}}]{Zohm_1999}%
  \BibitemOpen
  \bibfield  {author} {\bibinfo {author} {\bibfnamefont {H.}~\bibnamefont
  {Zohm}}, \bibinfo {author} {\bibfnamefont {G.}~\bibnamefont {Gantenbein}},
  \bibinfo {author} {\bibfnamefont {G.}~\bibnamefont {Giruzzi}}, \bibinfo
  {author} {\bibfnamefont {S.}~\bibnamefont {Günter}}, \bibinfo {author}
  {\bibfnamefont {F.}~\bibnamefont {Leuterer}}, \bibinfo {author}
  {\bibfnamefont {M.}~\bibnamefont {Maraschek}}, \bibinfo {author}
  {\bibfnamefont {J.}~\bibnamefont {Meskat}}, \bibinfo {author} {\bibfnamefont
  {A.}~\bibnamefont {Peeters}}, \bibinfo {author} {\bibfnamefont
  {W.}~\bibnamefont {Suttrop}}, \bibinfo {author} {\bibfnamefont
  {D.}~\bibnamefont {Wagner}}, \bibinfo {author} {\bibfnamefont
  {M.}~\bibnamefont {Zabi{\'{e}}go}}, \bibinfo {author} {\bibfnamefont {A.~U.}\
  \bibnamefont {Team}}, \ and\ \bibinfo {author} {\bibfnamefont
  {E.}~\bibnamefont {Group}},\ }\bibfield  {title} {\enquote {\bibinfo {title}
  {Experiments on neoclassical tearing mode stabilization by {ECCD} in {ASDEX}
  upgrade},}\ }\href {\doibase 10.1088/0029-5515/39/5/101} {\bibfield
  {journal} {\bibinfo  {journal} {Nuclear Fusion}\ }\textbf {\bibinfo {volume}
  {39}},\ \bibinfo {pages} {577--580} (\bibinfo {year} {1999})}\BibitemShut
  {NoStop}%
\bibitem [{\citenamefont {Prater}\ \emph {et~al.}(2003)\citenamefont {Prater},
  \citenamefont {Haye}, \citenamefont {Lohr}, \citenamefont {Luce},
  \citenamefont {Petty}, \citenamefont {Ferron}, \citenamefont {Humphreys},
  \citenamefont {Strait}, \citenamefont {Perkins},\ and\ \citenamefont
  {Harvey}}]{Prater_2003}%
  \BibitemOpen
  \bibfield  {author} {\bibinfo {author} {\bibfnamefont {R.}~\bibnamefont
  {Prater}}, \bibinfo {author} {\bibfnamefont {R.~L.}\ \bibnamefont {Haye}},
  \bibinfo {author} {\bibfnamefont {J.}~\bibnamefont {Lohr}}, \bibinfo {author}
  {\bibfnamefont {T.}~\bibnamefont {Luce}}, \bibinfo {author} {\bibfnamefont
  {C.}~\bibnamefont {Petty}}, \bibinfo {author} {\bibfnamefont
  {J.}~\bibnamefont {Ferron}}, \bibinfo {author} {\bibfnamefont
  {D.}~\bibnamefont {Humphreys}}, \bibinfo {author} {\bibfnamefont
  {E.}~\bibnamefont {Strait}}, \bibinfo {author} {\bibfnamefont
  {F.}~\bibnamefont {Perkins}}, \ and\ \bibinfo {author} {\bibfnamefont
  {R.}~\bibnamefont {Harvey}},\ }\bibfield  {title} {\enquote {\bibinfo {title}
  {Discharge improvement through control of neoclassical tearing modes by
  localized {ECCD} in {DIII}-d},}\ }\href {\doibase
  10.1088/0029-5515/43/10/014} {\bibfield  {journal} {\bibinfo  {journal}
  {Nuclear Fusion}\ }\textbf {\bibinfo {volume} {43}},\ \bibinfo {pages}
  {1128--1134} (\bibinfo {year} {2003})}\BibitemShut {NoStop}%
\bibitem [{\citenamefont {Warrick}\ \emph {et~al.}(2000)\citenamefont
  {Warrick}, \citenamefont {Buttery}, \citenamefont {Cunningham}, \citenamefont
  {Fielding}, \citenamefont {Hender}, \citenamefont {Lloyd}, \citenamefont
  {Morris}, \citenamefont {O'Brien}, \citenamefont {Pinfold}, \citenamefont
  {Stammers}, \citenamefont {Valovic}, \citenamefont {Walsh}, \citenamefont
  {Wilson}, \citenamefont {COMPASS-D},\ and\ \citenamefont
  {teams}}]{Warrick_2000}%
  \BibitemOpen
  \bibfield  {author} {\bibinfo {author} {\bibfnamefont {C.~D.}\ \bibnamefont
  {Warrick}}, \bibinfo {author} {\bibfnamefont {R.~J.}\ \bibnamefont
  {Buttery}}, \bibinfo {author} {\bibfnamefont {G.}~\bibnamefont {Cunningham}},
  \bibinfo {author} {\bibfnamefont {S.~J.}\ \bibnamefont {Fielding}}, \bibinfo
  {author} {\bibfnamefont {T.~C.}\ \bibnamefont {Hender}}, \bibinfo {author}
  {\bibfnamefont {B.}~\bibnamefont {Lloyd}}, \bibinfo {author} {\bibfnamefont
  {A.~W.}\ \bibnamefont {Morris}}, \bibinfo {author} {\bibfnamefont {M.~R.}\
  \bibnamefont {O'Brien}}, \bibinfo {author} {\bibfnamefont {T.}~\bibnamefont
  {Pinfold}}, \bibinfo {author} {\bibfnamefont {K.}~\bibnamefont {Stammers}},
  \bibinfo {author} {\bibfnamefont {M.}~\bibnamefont {Valovic}}, \bibinfo
  {author} {\bibfnamefont {M.}~\bibnamefont {Walsh}}, \bibinfo {author}
  {\bibfnamefont {H.~R.}\ \bibnamefont {Wilson}}, \bibinfo {author}
  {\bibnamefont {COMPASS-D}}, \ and\ \bibinfo {author} {\bibfnamefont
  {R.}~\bibnamefont {teams}},\ }\bibfield  {title} {\enquote {\bibinfo {title}
  {Complete stabilization of neoclassical tearing modes with lower hybrid
  current drive on compass-d},}\ }\href {\doibase 10.1103/PhysRevLett.85.574}
  {\bibfield  {journal} {\bibinfo  {journal} {Phys. Rev. Lett.}\ }\textbf
  {\bibinfo {volume} {85}},\ \bibinfo {pages} {574--577} (\bibinfo {year}
  {2000})}\BibitemShut {NoStop}%
\bibitem [{\citenamefont {Gantenbein}\ \emph {et~al.}(2000)\citenamefont
  {Gantenbein}, \citenamefont {Zohm}, \citenamefont {Giruzzi}, \citenamefont
  {G\"unter}, \citenamefont {Leuterer}, \citenamefont {Maraschek},
  \citenamefont {Meskat}, \citenamefont {Yu}, \citenamefont {Team},\ and\
  \citenamefont {(AUG)}}]{Gantenbein_2000}%
  \BibitemOpen
  \bibfield  {author} {\bibinfo {author} {\bibfnamefont {G.}~\bibnamefont
  {Gantenbein}}, \bibinfo {author} {\bibfnamefont {H.}~\bibnamefont {Zohm}},
  \bibinfo {author} {\bibfnamefont {G.}~\bibnamefont {Giruzzi}}, \bibinfo
  {author} {\bibfnamefont {S.}~\bibnamefont {G\"unter}}, \bibinfo {author}
  {\bibfnamefont {F.}~\bibnamefont {Leuterer}}, \bibinfo {author}
  {\bibfnamefont {M.}~\bibnamefont {Maraschek}}, \bibinfo {author}
  {\bibfnamefont {J.}~\bibnamefont {Meskat}}, \bibinfo {author} {\bibfnamefont
  {Q.}~\bibnamefont {Yu}}, \bibinfo {author} {\bibfnamefont {A.~U.}\
  \bibnamefont {Team}}, \ and\ \bibinfo {author} {\bibfnamefont {E.-G.}\
  \bibnamefont {(AUG)}},\ }\bibfield  {title} {\enquote {\bibinfo {title}
  {Complete suppression of neoclassical tearing modes with current drive at the
  electron-cyclotron-resonance frequency in asdex upgrade tokamak},}\ }\href
  {\doibase 10.1103/PhysRevLett.85.1242} {\bibfield  {journal} {\bibinfo
  {journal} {Phys. Rev. Lett.}\ }\textbf {\bibinfo {volume} {85}},\ \bibinfo
  {pages} {1242--1245} (\bibinfo {year} {2000})}\BibitemShut {NoStop}%
\bibitem [{\citenamefont {Zohm}\ \emph {et~al.}(2001)\citenamefont {Zohm},
  \citenamefont {Gantenbein}, \citenamefont {Gude}, \citenamefont {Günter},
  \citenamefont {Leuterer}, \citenamefont {Maraschek}, \citenamefont {Meskat},
  \citenamefont {Suttrop}, \citenamefont {Yu}, \citenamefont {Team},\ and\
  \citenamefont {(AUG)}}]{Zohm_2001}%
  \BibitemOpen
  \bibfield  {author} {\bibinfo {author} {\bibfnamefont {H.}~\bibnamefont
  {Zohm}}, \bibinfo {author} {\bibfnamefont {G.}~\bibnamefont {Gantenbein}},
  \bibinfo {author} {\bibfnamefont {A.}~\bibnamefont {Gude}}, \bibinfo {author}
  {\bibfnamefont {S.}~\bibnamefont {Günter}}, \bibinfo {author} {\bibfnamefont
  {F.}~\bibnamefont {Leuterer}}, \bibinfo {author} {\bibfnamefont
  {M.}~\bibnamefont {Maraschek}}, \bibinfo {author} {\bibfnamefont
  {J.}~\bibnamefont {Meskat}}, \bibinfo {author} {\bibfnamefont
  {W.}~\bibnamefont {Suttrop}}, \bibinfo {author} {\bibfnamefont
  {Q.}~\bibnamefont {Yu}}, \bibinfo {author} {\bibfnamefont {A.~U.}\
  \bibnamefont {Team}}, \ and\ \bibinfo {author} {\bibfnamefont {E.~G.}\
  \bibnamefont {(AUG)}},\ }\bibfield  {title} {\enquote {\bibinfo {title} {The
  physics of neoclassical tearing modes and their stabilization by {ECCD} in
  {ASDEX} upgrade},}\ }\href {\doibase 10.1088/0029-5515/41/2/306} {\bibfield
  {journal} {\bibinfo  {journal} {Nuclear Fusion}\ }\textbf {\bibinfo {volume}
  {41}},\ \bibinfo {pages} {197--202} (\bibinfo {year} {2001})}\BibitemShut
  {NoStop}%
\bibitem [{\citenamefont {Isayama}\ \emph {et~al.}(2000)\citenamefont
  {Isayama}, \citenamefont {Kamada}, \citenamefont {Ide}, \citenamefont
  {Hamamatsu}, \citenamefont {Oikawa}, \citenamefont {Suzuki}, \citenamefont
  {Neyatani}, \citenamefont {Ozeki}, \citenamefont {Ikeda},\ and\ \citenamefont
  {and}}]{Isayama_2000}%
  \BibitemOpen
  \bibfield  {author} {\bibinfo {author} {\bibfnamefont {A.}~\bibnamefont
  {Isayama}}, \bibinfo {author} {\bibfnamefont {Y.}~\bibnamefont {Kamada}},
  \bibinfo {author} {\bibfnamefont {S.}~\bibnamefont {Ide}}, \bibinfo {author}
  {\bibfnamefont {K.}~\bibnamefont {Hamamatsu}}, \bibinfo {author}
  {\bibfnamefont {T.}~\bibnamefont {Oikawa}}, \bibinfo {author} {\bibfnamefont
  {T.}~\bibnamefont {Suzuki}}, \bibinfo {author} {\bibfnamefont
  {Y.}~\bibnamefont {Neyatani}}, \bibinfo {author} {\bibfnamefont
  {T.}~\bibnamefont {Ozeki}}, \bibinfo {author} {\bibfnamefont
  {Y.}~\bibnamefont {Ikeda}}, \ and\ \bibinfo {author} {\bibfnamefont {K.~K.}\
  \bibnamefont {and}},\ }\bibfield  {title} {\enquote {\bibinfo {title}
  {Complete stabilization of a tearing mode in steady state high- ph-mode
  discharges by the first harmonic electron cyclotron heating/current drive on
  jt-60u},}\ }\href {\doibase 10.1088/0741-3335/42/12/102} {\bibfield
  {journal} {\bibinfo  {journal} {Plasma Physics and Controlled Fusion}\
  }\textbf {\bibinfo {volume} {42}},\ \bibinfo {pages} {L37--L45} (\bibinfo
  {year} {2000})}\BibitemShut {NoStop}%
\bibitem [{\citenamefont {La~Haye}\ \emph {et~al.}(2002)\citenamefont
  {La~Haye}, \citenamefont {Günter}, \citenamefont {Humphreys}, \citenamefont
  {Lohr}, \citenamefont {Luce}, \citenamefont {Maraschek}, \citenamefont
  {Petty}, \citenamefont {Prater}, \citenamefont {Scoville},\ and\
  \citenamefont {Strait}}]{La_Haye_2002}%
  \BibitemOpen
  \bibfield  {author} {\bibinfo {author} {\bibfnamefont {R.~J.}\ \bibnamefont
  {La~Haye}}, \bibinfo {author} {\bibfnamefont {S.}~\bibnamefont {Günter}},
  \bibinfo {author} {\bibfnamefont {D.~A.}\ \bibnamefont {Humphreys}}, \bibinfo
  {author} {\bibfnamefont {J.}~\bibnamefont {Lohr}}, \bibinfo {author}
  {\bibfnamefont {T.~C.}\ \bibnamefont {Luce}}, \bibinfo {author}
  {\bibfnamefont {M.~E.}\ \bibnamefont {Maraschek}}, \bibinfo {author}
  {\bibfnamefont {C.~C.}\ \bibnamefont {Petty}}, \bibinfo {author}
  {\bibfnamefont {R.}~\bibnamefont {Prater}}, \bibinfo {author} {\bibfnamefont
  {J.~T.}\ \bibnamefont {Scoville}}, \ and\ \bibinfo {author} {\bibfnamefont
  {E.~J.}\ \bibnamefont {Strait}},\ }\bibfield  {title} {\enquote {\bibinfo
  {title} {Control of neoclassical tearing modes in diii–d},}\ }\href
  {\doibase 10.1063/1.1456066} {\bibfield  {journal} {\bibinfo  {journal}
  {Physics of Plasmas}\ }\textbf {\bibinfo {volume} {9}},\ \bibinfo {pages}
  {2051--2060} (\bibinfo {year} {2002})}\BibitemShut {NoStop}%
\bibitem [{\citenamefont {Petty}\ \emph {et~al.}(2004)\citenamefont {Petty},
  \citenamefont {Haye}, \citenamefont {Luce}, \citenamefont {Humphreys},
  \citenamefont {Hyatt}, \citenamefont {Lohr}, \citenamefont {Prater},
  \citenamefont {Strait},\ and\ \citenamefont {Wade}}]{Petty_2004}%
  \BibitemOpen
  \bibfield  {author} {\bibinfo {author} {\bibfnamefont {C.}~\bibnamefont
  {Petty}}, \bibinfo {author} {\bibfnamefont {R.~L.}\ \bibnamefont {Haye}},
  \bibinfo {author} {\bibfnamefont {T.}~\bibnamefont {Luce}}, \bibinfo {author}
  {\bibfnamefont {D.}~\bibnamefont {Humphreys}}, \bibinfo {author}
  {\bibfnamefont {A.}~\bibnamefont {Hyatt}}, \bibinfo {author} {\bibfnamefont
  {J.}~\bibnamefont {Lohr}}, \bibinfo {author} {\bibfnamefont {R.}~\bibnamefont
  {Prater}}, \bibinfo {author} {\bibfnamefont {E.}~\bibnamefont {Strait}}, \
  and\ \bibinfo {author} {\bibfnamefont {M.}~\bibnamefont {Wade}},\ }\bibfield
  {title} {\enquote {\bibinfo {title} {Complete suppression of them= 2/n= 1
  neoclassical tearing mode using electron cyclotron current drive in
  {DIII}-d},}\ }\href {\doibase 10.1088/0029-5515/44/2/004} {\bibfield
  {journal} {\bibinfo  {journal} {Nuclear Fusion}\ }\textbf {\bibinfo {volume}
  {44}},\ \bibinfo {pages} {243--251} (\bibinfo {year} {2004})}\BibitemShut
  {NoStop}%
\bibitem [{\citenamefont {Volpe}\ \emph {et~al.}(2015)\citenamefont {Volpe},
  \citenamefont {Hyatt}, \citenamefont {La~Haye}, \citenamefont {Lanctot},
  \citenamefont {Lohr}, \citenamefont {Prater}, \citenamefont {Strait},\ and\
  \citenamefont {Welander}}]{Volpe_2015}%
  \BibitemOpen
  \bibfield  {author} {\bibinfo {author} {\bibfnamefont {F.~A.}\ \bibnamefont
  {Volpe}}, \bibinfo {author} {\bibfnamefont {A.}~\bibnamefont {Hyatt}},
  \bibinfo {author} {\bibfnamefont {R.~J.}\ \bibnamefont {La~Haye}}, \bibinfo
  {author} {\bibfnamefont {M.~J.}\ \bibnamefont {Lanctot}}, \bibinfo {author}
  {\bibfnamefont {J.}~\bibnamefont {Lohr}}, \bibinfo {author} {\bibfnamefont
  {R.}~\bibnamefont {Prater}}, \bibinfo {author} {\bibfnamefont {E.~J.}\
  \bibnamefont {Strait}}, \ and\ \bibinfo {author} {\bibfnamefont
  {A.}~\bibnamefont {Welander}},\ }\bibfield  {title} {\enquote {\bibinfo
  {title} {Avoiding tokamak disruptions by applying static magnetic fields that
  align locked modes with stabilizing wave-driven currents},}\ }\href {\doibase
  10.1103/PhysRevLett.115.175002} {\bibfield  {journal} {\bibinfo  {journal}
  {Phys. Rev. Lett.}\ }\textbf {\bibinfo {volume} {115}},\ \bibinfo {pages}
  {175002} (\bibinfo {year} {2015})}\BibitemShut {NoStop}%
\bibitem [{\citenamefont {Nelson}\ \emph {et~al.}(2019)\citenamefont {Nelson},
  \citenamefont {Haye}, \citenamefont {Austin}, \citenamefont {Welander},\ and\
  \citenamefont {Kolemen}}]{Nelson_2019}%
  \BibitemOpen
  \bibfield  {author} {\bibinfo {author} {\bibfnamefont {A.}~\bibnamefont
  {Nelson}}, \bibinfo {author} {\bibfnamefont {R.~L.}\ \bibnamefont {Haye}},
  \bibinfo {author} {\bibfnamefont {M.}~\bibnamefont {Austin}}, \bibinfo
  {author} {\bibfnamefont {A.}~\bibnamefont {Welander}}, \ and\ \bibinfo
  {author} {\bibfnamefont {E.}~\bibnamefont {Kolemen}},\ }\bibfield  {title}
  {\enquote {\bibinfo {title} {Simultaneous detection of neoclassical tearing
  mode and electron cyclotron current drive locations using electron cyclotron
  emission in diii-d},}\ }\href {\doibase
  https://doi.org/10.1016/j.fusengdes.2019.02.089} {\bibfield  {journal}
  {\bibinfo  {journal} {Fusion Engineering and Design}\ }\textbf {\bibinfo
  {volume} {141}},\ \bibinfo {pages} {25 -- 29} (\bibinfo {year}
  {2019})}\BibitemShut {NoStop}%
\bibitem [{\citenamefont {Reiman}\ and\ \citenamefont
  {Fisch}(2018)}]{Reiman_2018}%
  \BibitemOpen
  \bibfield  {author} {\bibinfo {author} {\bibfnamefont {A.~H.}\ \bibnamefont
  {Reiman}}\ and\ \bibinfo {author} {\bibfnamefont {N.~J.}\ \bibnamefont
  {Fisch}},\ }\bibfield  {title} {\enquote {\bibinfo {title} {Suppression of
  tearing modes by radio frequency current condensation},}\ }\href {\doibase
  10.1103/PhysRevLett.121.225001} {\bibfield  {journal} {\bibinfo  {journal}
  {Phys. Rev. Lett.}\ }\textbf {\bibinfo {volume} {121}},\ \bibinfo {pages}
  {225001} (\bibinfo {year} {2018})}\BibitemShut {NoStop}%
\bibitem [{\citenamefont {Fisch}(1978)}]{fisch_1978}%
  \BibitemOpen
  \bibfield  {author} {\bibinfo {author} {\bibfnamefont {N.~J.}\ \bibnamefont
  {Fisch}},\ }\bibfield  {title} {\enquote {\bibinfo {title} {Confining a
  tokamak plasma with rf-driven currents},}\ }\href {\doibase
  10.1103/PhysRevLett.41.873} {\bibfield  {journal} {\bibinfo  {journal} {Phys.
  Rev. Lett.}\ }\textbf {\bibinfo {volume} {41}},\ \bibinfo {pages} {873--876}
  (\bibinfo {year} {1978})}\BibitemShut {NoStop}%
\bibitem [{\citenamefont {Fisch}\ and\ \citenamefont
  {Boozer}(1980)}]{fisch_1980}%
  \BibitemOpen
  \bibfield  {author} {\bibinfo {author} {\bibfnamefont {N.~J.}\ \bibnamefont
  {Fisch}}\ and\ \bibinfo {author} {\bibfnamefont {A.~H.}\ \bibnamefont
  {Boozer}},\ }\bibfield  {title} {\enquote {\bibinfo {title} {Creating an
  asymmetric plasma resistivity with waves},}\ }\href {\doibase
  10.1103/PhysRevLett.45.720} {\bibfield  {journal} {\bibinfo  {journal} {Phys.
  Rev. Lett.}\ }\textbf {\bibinfo {volume} {45}},\ \bibinfo {pages} {720--722}
  (\bibinfo {year} {1980})}\BibitemShut {NoStop}%
\bibitem [{\citenamefont {Fisch}(1987)}]{fisch_1987}%
  \BibitemOpen
  \bibfield  {author} {\bibinfo {author} {\bibfnamefont {N.~J.}\ \bibnamefont
  {Fisch}},\ }\bibfield  {title} {\enquote {\bibinfo {title} {Theory of current
  drive in plasmas},}\ }\href {\doibase 10.1103/RevModPhys.59.175} {\bibfield
  {journal} {\bibinfo  {journal} {Rev. Mod. Phys.}\ }\textbf {\bibinfo {volume}
  {59}},\ \bibinfo {pages} {175--234} (\bibinfo {year} {1987})}\BibitemShut
  {NoStop}%
\bibitem [{\citenamefont {Westerhof}\ \emph {et~al.}(2007)\citenamefont
  {Westerhof}, \citenamefont {Lazaros}, \citenamefont {Farshi}, \citenamefont
  {de~Baar}, \citenamefont {de~Bock}, \citenamefont {Classen}, \citenamefont
  {Jaspers}, \citenamefont {Hogeweij}, \citenamefont {Koslowski}, \citenamefont
  {Krämer-Flecken}, \citenamefont {Liang}, \citenamefont {Cardozo},\ and\
  \citenamefont {Zimmermann}}]{Westerhof_2007}%
  \BibitemOpen
  \bibfield  {author} {\bibinfo {author} {\bibfnamefont {E.}~\bibnamefont
  {Westerhof}}, \bibinfo {author} {\bibfnamefont {A.}~\bibnamefont {Lazaros}},
  \bibinfo {author} {\bibfnamefont {E.}~\bibnamefont {Farshi}}, \bibinfo
  {author} {\bibfnamefont {M.}~\bibnamefont {de~Baar}}, \bibinfo {author}
  {\bibfnamefont {M.}~\bibnamefont {de~Bock}}, \bibinfo {author} {\bibfnamefont
  {I.}~\bibnamefont {Classen}}, \bibinfo {author} {\bibfnamefont
  {R.}~\bibnamefont {Jaspers}}, \bibinfo {author} {\bibfnamefont
  {G.}~\bibnamefont {Hogeweij}}, \bibinfo {author} {\bibfnamefont
  {H.}~\bibnamefont {Koslowski}}, \bibinfo {author} {\bibfnamefont
  {A.}~\bibnamefont {Krämer-Flecken}}, \bibinfo {author} {\bibfnamefont
  {Y.}~\bibnamefont {Liang}}, \bibinfo {author} {\bibfnamefont {N.~L.}\
  \bibnamefont {Cardozo}}, \ and\ \bibinfo {author} {\bibfnamefont
  {O.}~\bibnamefont {Zimmermann}},\ }\bibfield  {title} {\enquote {\bibinfo
  {title} {Tearing mode stabilization by electron cyclotron resonance heating
  demonstrated in the {TEXTOR} tokamak and the implication for {ITER}},}\
  }\href {\doibase 10.1088/0029-5515/47/2/003} {\bibfield  {journal} {\bibinfo
  {journal} {Nuclear Fusion}\ }\textbf {\bibinfo {volume} {47}},\ \bibinfo
  {pages} {85--90} (\bibinfo {year} {2007})}\BibitemShut {NoStop}%
\bibitem [{\citenamefont {Rodríguez}, \citenamefont {Reiman},\ and\
  \citenamefont {Fisch}(2019)}]{Rodriguez_2019}%
  \BibitemOpen
  \bibfield  {author} {\bibinfo {author} {\bibfnamefont {E.}~\bibnamefont
  {Rodríguez}}, \bibinfo {author} {\bibfnamefont {A.~H.}\ \bibnamefont
  {Reiman}}, \ and\ \bibinfo {author} {\bibfnamefont {N.~J.}\ \bibnamefont
  {Fisch}},\ }\bibfield  {title} {\enquote {\bibinfo {title} {Rf current
  condensation in magnetic islands and associated hysteresis phenomena},}\
  }\href {\doibase 10.1063/1.5118424} {\bibfield  {journal} {\bibinfo
  {journal} {Physics of Plasmas}\ }\textbf {\bibinfo {volume} {26}},\ \bibinfo
  {pages} {092511} (\bibinfo {year} {2019})}\BibitemShut {NoStop}%
\bibitem [{\citenamefont {Bonoli}\ and\ \citenamefont
  {Englade}(1986)}]{Bonoli_1986}%
  \BibitemOpen
  \bibfield  {author} {\bibinfo {author} {\bibfnamefont {P.~T.}\ \bibnamefont
  {Bonoli}}\ and\ \bibinfo {author} {\bibfnamefont {R.~C.}\ \bibnamefont
  {Englade}},\ }\bibfield  {title} {\enquote {\bibinfo {title} {Simulation
  model for lower hybrid current drive},}\ }\href {\doibase 10.1063/1.865494}
  {\bibfield  {journal} {\bibinfo  {journal} {The Physics of Fluids}\ }\textbf
  {\bibinfo {volume} {29}},\ \bibinfo {pages} {2937--2950} (\bibinfo {year}
  {1986})}\BibitemShut {NoStop}%
\bibitem [{\citenamefont {Prater}\ \emph {et~al.}(2008)\citenamefont {Prater},
  \citenamefont {Farina}, \citenamefont {Gribov}, \citenamefont {Harvey},
  \citenamefont {Ram}, \citenamefont {Lin-Liu}, \citenamefont {Poli},
  \citenamefont {Smirnov}, \citenamefont {Volpe}, \citenamefont {Westerhof},\
  and\ \citenamefont {and}}]{Prater_2008}%
  \BibitemOpen
  \bibfield  {author} {\bibinfo {author} {\bibfnamefont {R.}~\bibnamefont
  {Prater}}, \bibinfo {author} {\bibfnamefont {D.}~\bibnamefont {Farina}},
  \bibinfo {author} {\bibfnamefont {Y.}~\bibnamefont {Gribov}}, \bibinfo
  {author} {\bibfnamefont {R.}~\bibnamefont {Harvey}}, \bibinfo {author}
  {\bibfnamefont {A.}~\bibnamefont {Ram}}, \bibinfo {author} {\bibfnamefont
  {Y.}~\bibnamefont {Lin-Liu}}, \bibinfo {author} {\bibfnamefont
  {E.}~\bibnamefont {Poli}}, \bibinfo {author} {\bibfnamefont {A.}~\bibnamefont
  {Smirnov}}, \bibinfo {author} {\bibfnamefont {F.}~\bibnamefont {Volpe}},
  \bibinfo {author} {\bibfnamefont {E.}~\bibnamefont {Westerhof}}, \ and\
  \bibinfo {author} {\bibfnamefont {A.~Z.}\ \bibnamefont {and}},\ }\bibfield
  {title} {\enquote {\bibinfo {title} {Benchmarking of codes for electron
  cyclotron heating and electron cyclotron current drive under {ITER}
  conditions},}\ }\href {\doibase 10.1088/0029-5515/48/3/035006} {\bibfield
  {journal} {\bibinfo  {journal} {Nuclear Fusion}\ }\textbf {\bibinfo {volume}
  {48}},\ \bibinfo {pages} {035006} (\bibinfo {year} {2008})}\BibitemShut
  {NoStop}%
\bibitem [{\citenamefont {Kurita}\ \emph {et~al.}(1994)\citenamefont {Kurita},
  \citenamefont {Tuda}, \citenamefont {Azumi}, \citenamefont {Takizuka},\ and\
  \citenamefont {Takeda}}]{Kurita_1994}%
  \BibitemOpen
  \bibfield  {author} {\bibinfo {author} {\bibfnamefont {G.}~\bibnamefont
  {Kurita}}, \bibinfo {author} {\bibfnamefont {T.}~\bibnamefont {Tuda}},
  \bibinfo {author} {\bibfnamefont {M.}~\bibnamefont {Azumi}}, \bibinfo
  {author} {\bibfnamefont {T.}~\bibnamefont {Takizuka}}, \ and\ \bibinfo
  {author} {\bibfnamefont {T.}~\bibnamefont {Takeda}},\ }\bibfield  {title}
  {\enquote {\bibinfo {title} {Effect of local heating on the m=2 tearing mode
  in a tokamak},}\ }\href {\doibase 10.1088/0029-5515/34/11/i08} {\bibfield
  {journal} {\bibinfo  {journal} {Nuclear Fusion}\ }\textbf {\bibinfo {volume}
  {34}},\ \bibinfo {pages} {1497--1515} (\bibinfo {year} {1994})}\BibitemShut
  {NoStop}%
\bibitem [{\citenamefont {Hegna}\ and\ \citenamefont
  {Callen}(1997)}]{Hegna_1997}%
  \BibitemOpen
  \bibfield  {author} {\bibinfo {author} {\bibfnamefont {C.~C.}\ \bibnamefont
  {Hegna}}\ and\ \bibinfo {author} {\bibfnamefont {J.~D.}\ \bibnamefont
  {Callen}},\ }\bibfield  {title} {\enquote {\bibinfo {title} {On the
  stabilization of neoclassical magnetohydrodynamic tearing modes using
  localized current drive or heating},}\ }\href {\doibase 10.1063/1.872426}
  {\bibfield  {journal} {\bibinfo  {journal} {Physics of Plasmas}\ }\textbf
  {\bibinfo {volume} {4}},\ \bibinfo {pages} {2940--2946} (\bibinfo {year}
  {1997})}\BibitemShut {NoStop}%
\bibitem [{\citenamefont {Lazzari}\ and\ \citenamefont
  {Westerhof}(2009)}]{De_Lazzari_2009}%
  \BibitemOpen
  \bibfield  {author} {\bibinfo {author} {\bibfnamefont {D.~D.}\ \bibnamefont
  {Lazzari}}\ and\ \bibinfo {author} {\bibfnamefont {E.}~\bibnamefont
  {Westerhof}},\ }\bibfield  {title} {\enquote {\bibinfo {title} {On the merits
  of heating and current drive for tearing mode stabilization},}\ }\href
  {\doibase 10.1088/0029-5515/49/7/075002} {\bibfield  {journal} {\bibinfo
  {journal} {Nuclear Fusion}\ }\textbf {\bibinfo {volume} {49}},\ \bibinfo
  {pages} {075002} (\bibinfo {year} {2009})}\BibitemShut {NoStop}%
\bibitem [{\citenamefont {Rodriguez}, \citenamefont {Reiman},\ and\
  \citenamefont {Fisch}(2020)}]{eduardo}%
  \BibitemOpen
  \bibfield  {author} {\bibinfo {author} {\bibfnamefont {E.}~\bibnamefont
  {Rodriguez}}, \bibinfo {author} {\bibfnamefont {A.~H.}\ \bibnamefont
  {Reiman}}, \ and\ \bibinfo {author} {\bibfnamefont {N.~J.}\ \bibnamefont
  {Fisch}},\ }\href@noop {} {\enquote {\bibinfo {title} {Rf current
  condensation in the presence of turbulent enhanced transport},}\ } (\bibinfo
  {year} {2020}),\ \Eprint {http://arxiv.org/abs/2001.09044} {arXiv:2001.09044}
  \BibitemShut {NoStop}%
\bibitem [{\citenamefont {Maraschek}\ \emph {et~al.}(2007)\citenamefont
  {Maraschek}, \citenamefont {Gantenbein}, \citenamefont {Yu}, \citenamefont
  {Zohm}, \citenamefont {G\"unter}, \citenamefont {Leuterer},\ and\
  \citenamefont {Manini}}]{Maraschek_2007}%
  \BibitemOpen
  \bibfield  {author} {\bibinfo {author} {\bibfnamefont {M.}~\bibnamefont
  {Maraschek}}, \bibinfo {author} {\bibfnamefont {G.}~\bibnamefont
  {Gantenbein}}, \bibinfo {author} {\bibfnamefont {Q.}~\bibnamefont {Yu}},
  \bibinfo {author} {\bibfnamefont {H.}~\bibnamefont {Zohm}}, \bibinfo {author}
  {\bibfnamefont {S.}~\bibnamefont {G\"unter}}, \bibinfo {author}
  {\bibfnamefont {F.}~\bibnamefont {Leuterer}}, \ and\ \bibinfo {author}
  {\bibfnamefont {A.}~\bibnamefont {Manini}} (\bibinfo {collaboration} {ECRH
  Group and ASDEX Upgrade Team}),\ }\bibfield  {title} {\enquote {\bibinfo
  {title} {Enhancement of the stabilization efficiency of a neoclassical
  magnetic island by modulated electron cyclotron current drive in the asdex
  upgrade tokamak},}\ }\href {\doibase 10.1103/PhysRevLett.98.025005}
  {\bibfield  {journal} {\bibinfo  {journal} {Phys. Rev. Lett.}\ }\textbf
  {\bibinfo {volume} {98}},\ \bibinfo {pages} {025005} (\bibinfo {year}
  {2007})}\BibitemShut {NoStop}%
\bibitem [{\citenamefont {Volpe}\ \emph {et~al.}(2009)\citenamefont {Volpe},
  \citenamefont {Austin}, \citenamefont {La~Haye}, \citenamefont {Lohr},
  \citenamefont {Prater}, \citenamefont {Strait},\ and\ \citenamefont
  {Welander}}]{Volpe_2009}%
  \BibitemOpen
  \bibfield  {author} {\bibinfo {author} {\bibfnamefont {F.~A.~G.}\
  \bibnamefont {Volpe}}, \bibinfo {author} {\bibfnamefont {M.~E.}\ \bibnamefont
  {Austin}}, \bibinfo {author} {\bibfnamefont {R.~J.}\ \bibnamefont {La~Haye}},
  \bibinfo {author} {\bibfnamefont {J.}~\bibnamefont {Lohr}}, \bibinfo {author}
  {\bibfnamefont {R.}~\bibnamefont {Prater}}, \bibinfo {author} {\bibfnamefont
  {E.~J.}\ \bibnamefont {Strait}}, \ and\ \bibinfo {author} {\bibfnamefont
  {A.~S.}\ \bibnamefont {Welander}},\ }\bibfield  {title} {\enquote {\bibinfo
  {title} {Advanced techniques for neoclassical tearing mode control in
  diii-d},}\ }\href {\doibase 10.1063/1.3232325} {\bibfield  {journal}
  {\bibinfo  {journal} {Physics of Plasmas}\ }\textbf {\bibinfo {volume}
  {16}},\ \bibinfo {pages} {102502} (\bibinfo {year} {2009})}\BibitemShut
  {NoStop}%
\bibitem [{\citenamefont {Isayama}\ \emph {et~al.}(2009)\citenamefont
  {Isayama}, \citenamefont {Matsunaga}, \citenamefont {Kobayashi},
  \citenamefont {Moriyama}, \citenamefont {Oyama}, \citenamefont {Sakamoto},
  \citenamefont {Suzuki}, \citenamefont {Urano}, \citenamefont {Hayashi},
  \citenamefont {Kamada}, \citenamefont {Ozeki}, \citenamefont {Hirano},
  \citenamefont {Urso}, \citenamefont {Zohm}, \citenamefont {Maraschek},
  \citenamefont {Hobirk},\ and\ \citenamefont {and}}]{Isayama_2009}%
  \BibitemOpen
  \bibfield  {author} {\bibinfo {author} {\bibfnamefont {A.}~\bibnamefont
  {Isayama}}, \bibinfo {author} {\bibfnamefont {G.}~\bibnamefont {Matsunaga}},
  \bibinfo {author} {\bibfnamefont {T.}~\bibnamefont {Kobayashi}}, \bibinfo
  {author} {\bibfnamefont {S.}~\bibnamefont {Moriyama}}, \bibinfo {author}
  {\bibfnamefont {N.}~\bibnamefont {Oyama}}, \bibinfo {author} {\bibfnamefont
  {Y.}~\bibnamefont {Sakamoto}}, \bibinfo {author} {\bibfnamefont
  {T.}~\bibnamefont {Suzuki}}, \bibinfo {author} {\bibfnamefont
  {H.}~\bibnamefont {Urano}}, \bibinfo {author} {\bibfnamefont
  {N.}~\bibnamefont {Hayashi}}, \bibinfo {author} {\bibfnamefont
  {Y.}~\bibnamefont {Kamada}}, \bibinfo {author} {\bibfnamefont
  {T.}~\bibnamefont {Ozeki}}, \bibinfo {author} {\bibfnamefont
  {Y.}~\bibnamefont {Hirano}}, \bibinfo {author} {\bibfnamefont
  {L.}~\bibnamefont {Urso}}, \bibinfo {author} {\bibfnamefont {H.}~\bibnamefont
  {Zohm}}, \bibinfo {author} {\bibfnamefont {M.}~\bibnamefont {Maraschek}},
  \bibinfo {author} {\bibfnamefont {J.}~\bibnamefont {Hobirk}}, \ and\ \bibinfo
  {author} {\bibfnamefont {K.~N.}\ \bibnamefont {and}},\ }\bibfield  {title}
  {\enquote {\bibinfo {title} {Neoclassical tearing mode control using electron
  cyclotron current drive and magnetic island evolution in {JT}-60u},}\ }\href
  {\doibase 10.1088/0029-5515/49/5/055006} {\bibfield  {journal} {\bibinfo
  {journal} {Nuclear Fusion}\ }\textbf {\bibinfo {volume} {49}},\ \bibinfo
  {pages} {055006} (\bibinfo {year} {2009})}\BibitemShut {NoStop}%
\bibitem [{\citenamefont {Kasparek}\ \emph {et~al.}(2016)\citenamefont
  {Kasparek}, \citenamefont {Doelman}, \citenamefont {Stober}, \citenamefont
  {Maraschek}, \citenamefont {Zohm}, \citenamefont {Monaco}, \citenamefont
  {Eixenberger}, \citenamefont {Klop}, \citenamefont {Wagner}, \citenamefont
  {Schubert}, \citenamefont {Schütz}, \citenamefont {Grünwald}, \citenamefont
  {Plaum}, \citenamefont {Munk},\ and\ \citenamefont {and}}]{Kasparek_2016}%
  \BibitemOpen
  \bibfield  {author} {\bibinfo {author} {\bibfnamefont {W.}~\bibnamefont
  {Kasparek}}, \bibinfo {author} {\bibfnamefont {N.}~\bibnamefont {Doelman}},
  \bibinfo {author} {\bibfnamefont {J.}~\bibnamefont {Stober}}, \bibinfo
  {author} {\bibfnamefont {M.}~\bibnamefont {Maraschek}}, \bibinfo {author}
  {\bibfnamefont {H.}~\bibnamefont {Zohm}}, \bibinfo {author} {\bibfnamefont
  {F.}~\bibnamefont {Monaco}}, \bibinfo {author} {\bibfnamefont
  {H.}~\bibnamefont {Eixenberger}}, \bibinfo {author} {\bibfnamefont
  {W.}~\bibnamefont {Klop}}, \bibinfo {author} {\bibfnamefont {D.}~\bibnamefont
  {Wagner}}, \bibinfo {author} {\bibfnamefont {M.}~\bibnamefont {Schubert}},
  \bibinfo {author} {\bibfnamefont {H.}~\bibnamefont {Schütz}}, \bibinfo
  {author} {\bibfnamefont {G.}~\bibnamefont {Grünwald}}, \bibinfo {author}
  {\bibfnamefont {B.}~\bibnamefont {Plaum}}, \bibinfo {author} {\bibfnamefont
  {R.}~\bibnamefont {Munk}}, \ and\ \bibinfo {author} {\bibfnamefont {K.~S.}\
  \bibnamefont {and}},\ }\bibfield  {title} {\enquote {\bibinfo {title} {{NTM}
  stabilization by alternating o-point {EC} current drive using a high-power
  diplexer},}\ }\href {\doibase 10.1088/0029-5515/56/12/126001} {\bibfield
  {journal} {\bibinfo  {journal} {Nuclear Fusion}\ }\textbf {\bibinfo {volume}
  {56}},\ \bibinfo {pages} {126001} (\bibinfo {year} {2016})}\BibitemShut
  {NoStop}%
\bibitem [{\citenamefont {Zohm}\ \emph {et~al.}(2007)\citenamefont {Zohm},
  \citenamefont {Gantenbein}, \citenamefont {Leuterer}, \citenamefont {Manini},
  \citenamefont {Maraschek}, \citenamefont {Yu},\ and\ \citenamefont {the ASDEX
  Upgrade~Team}}]{Zohm_2007}%
  \BibitemOpen
  \bibfield  {author} {\bibinfo {author} {\bibfnamefont {H.}~\bibnamefont
  {Zohm}}, \bibinfo {author} {\bibfnamefont {G.}~\bibnamefont {Gantenbein}},
  \bibinfo {author} {\bibfnamefont {F.}~\bibnamefont {Leuterer}}, \bibinfo
  {author} {\bibfnamefont {A.}~\bibnamefont {Manini}}, \bibinfo {author}
  {\bibfnamefont {M.}~\bibnamefont {Maraschek}}, \bibinfo {author}
  {\bibfnamefont {Q.}~\bibnamefont {Yu}}, \ and\ \bibinfo {author}
  {\bibnamefont {the ASDEX Upgrade~Team}},\ }\bibfield  {title} {\enquote
  {\bibinfo {title} {Control of {MHD} instabilities by {ECCD}: {ASDEX} upgrade
  results and implications for {ITER}},}\ }\href {\doibase
  10.1088/0029-5515/47/3/010} {\bibfield  {journal} {\bibinfo  {journal}
  {Nuclear Fusion}\ }\textbf {\bibinfo {volume} {47}},\ \bibinfo {pages}
  {228--232} (\bibinfo {year} {2007})}\BibitemShut {NoStop}%
\bibitem [{\citenamefont {Erba}\ \emph {et~al.}(1998)\citenamefont {Erba},
  \citenamefont {Aniel}, \citenamefont {Basiuk}, \citenamefont {Becoulet},\
  and\ \citenamefont {Litaudon}}]{Erba_1998}%
  \BibitemOpen
  \bibfield  {author} {\bibinfo {author} {\bibfnamefont {M.}~\bibnamefont
  {Erba}}, \bibinfo {author} {\bibfnamefont {T.}~\bibnamefont {Aniel}},
  \bibinfo {author} {\bibfnamefont {V.}~\bibnamefont {Basiuk}}, \bibinfo
  {author} {\bibfnamefont {A.}~\bibnamefont {Becoulet}}, \ and\ \bibinfo
  {author} {\bibfnamefont {X.}~\bibnamefont {Litaudon}},\ }\bibfield  {title}
  {\enquote {\bibinfo {title} {Validation of a new mixed bohm/gyro-bohm model
  for electron and ion heat transport against the {ITER}, tore supra and
  {START} database discharges},}\ }\href {\doibase 10.1088/0029-5515/38/7/305}
  {\bibfield  {journal} {\bibinfo  {journal} {Nuclear Fusion}\ }\textbf
  {\bibinfo {volume} {38}},\ \bibinfo {pages} {1013--1028} (\bibinfo {year}
  {1998})}\BibitemShut {NoStop}%
\bibitem [{\citenamefont {Poli}\ \emph {et~al.}(2017)\citenamefont {Poli},
  \citenamefont {Fredrickson}, \citenamefont {Henderson}, \citenamefont {Kim},
  \citenamefont {Bertelli}, \citenamefont {Poli}, \citenamefont {Farina},\ and\
  \citenamefont {Figini}}]{Poli_2017}%
  \BibitemOpen
  \bibfield  {author} {\bibinfo {author} {\bibfnamefont {F.}~\bibnamefont
  {Poli}}, \bibinfo {author} {\bibfnamefont {E.}~\bibnamefont {Fredrickson}},
  \bibinfo {author} {\bibfnamefont {M.}~\bibnamefont {Henderson}}, \bibinfo
  {author} {\bibfnamefont {S.-H.}\ \bibnamefont {Kim}}, \bibinfo {author}
  {\bibfnamefont {N.}~\bibnamefont {Bertelli}}, \bibinfo {author}
  {\bibfnamefont {E.}~\bibnamefont {Poli}}, \bibinfo {author} {\bibfnamefont
  {D.}~\bibnamefont {Farina}}, \ and\ \bibinfo {author} {\bibfnamefont
  {L.}~\bibnamefont {Figini}},\ }\bibfield  {title} {\enquote {\bibinfo {title}
  {Electron cyclotron power management for control of neoclassical tearing
  modes in the {ITER} baseline scenario},}\ }\href {\doibase
  10.1088/1741-4326/aa8e0b} {\bibfield  {journal} {\bibinfo  {journal} {Nuclear
  Fusion}\ }\textbf {\bibinfo {volume} {58}},\ \bibinfo {pages} {016007}
  (\bibinfo {year} {2017})}\BibitemShut {NoStop}%
\bibitem [{\citenamefont {Yoshioka}, \citenamefont {Kinoshha},\ and\
  \citenamefont {Kobayashi}(1984)}]{Yoshioka_1984}%
  \BibitemOpen
  \bibfield  {author} {\bibinfo {author} {\bibfnamefont {Y.}~\bibnamefont
  {Yoshioka}}, \bibinfo {author} {\bibfnamefont {S.}~\bibnamefont {Kinoshha}},
  \ and\ \bibinfo {author} {\bibfnamefont {T.}~\bibnamefont {Kobayashi}},\
  }\bibfield  {title} {\enquote {\bibinfo {title} {Numerical study of magnetic
  island suppression by {RF} waves in large tokamaks},}\ }\href {\doibase
  10.1088/0029-5515/24/5/004} {\bibfield  {journal} {\bibinfo  {journal}
  {Nuclear Fusion}\ }\textbf {\bibinfo {volume} {24}},\ \bibinfo {pages}
  {565--572} (\bibinfo {year} {1984})}\BibitemShut {NoStop}%
\bibitem [{\citenamefont {Yu}\ \emph {et~al.}(2000)\citenamefont {Yu},
  \citenamefont {Günter}, \citenamefont {Giruzzi}, \citenamefont {Lackner},\
  and\ \citenamefont {Zabiego}}]{Yu_2000}%
  \BibitemOpen
  \bibfield  {author} {\bibinfo {author} {\bibfnamefont {Q.}~\bibnamefont
  {Yu}}, \bibinfo {author} {\bibfnamefont {S.}~\bibnamefont {Günter}},
  \bibinfo {author} {\bibfnamefont {G.}~\bibnamefont {Giruzzi}}, \bibinfo
  {author} {\bibfnamefont {K.}~\bibnamefont {Lackner}}, \ and\ \bibinfo
  {author} {\bibfnamefont {M.}~\bibnamefont {Zabiego}},\ }\bibfield  {title}
  {\enquote {\bibinfo {title} {Modeling of the stabilization of neoclassical
  tearing modes by localized radio frequency current drive},}\ }\href {\doibase
  10.1063/1.873799} {\bibfield  {journal} {\bibinfo  {journal} {Physics of
  Plasmas}\ }\textbf {\bibinfo {volume} {7}},\ \bibinfo {pages} {312--322}
  (\bibinfo {year} {2000})}\BibitemShut {NoStop}%
\bibitem [{\citenamefont {Bertelli}, \citenamefont {Lazzari},\ and\
  \citenamefont {Westerhof}(2011)}]{Bertelli_2011}%
  \BibitemOpen
  \bibfield  {author} {\bibinfo {author} {\bibfnamefont {N.}~\bibnamefont
  {Bertelli}}, \bibinfo {author} {\bibfnamefont {D.~D.}\ \bibnamefont
  {Lazzari}}, \ and\ \bibinfo {author} {\bibfnamefont {E.}~\bibnamefont
  {Westerhof}},\ }\bibfield  {title} {\enquote {\bibinfo {title} {Requirements
  on localized current drive for the suppression of neoclassical tearing
  modes},}\ }\href {\doibase 10.1088/0029-5515/51/10/103007} {\bibfield
  {journal} {\bibinfo  {journal} {Nuclear Fusion}\ }\textbf {\bibinfo {volume}
  {51}},\ \bibinfo {pages} {103007} (\bibinfo {year} {2011})}\BibitemShut
  {NoStop}%
\bibitem [{\citenamefont {Poli}\ \emph {et~al.}(2015)\citenamefont {Poli},
  \citenamefont {Angioni}, \citenamefont {Casson}, \citenamefont {Farina},
  \citenamefont {Figini}, \citenamefont {Goodman}, \citenamefont {Maj},
  \citenamefont {Sauter}, \citenamefont {Weber}, \citenamefont {Zohm},
  \citenamefont {Saibene},\ and\ \citenamefont {Henderson}}]{Poli_2015}%
  \BibitemOpen
  \bibfield  {author} {\bibinfo {author} {\bibfnamefont {E.}~\bibnamefont
  {Poli}}, \bibinfo {author} {\bibfnamefont {C.}~\bibnamefont {Angioni}},
  \bibinfo {author} {\bibfnamefont {F.}~\bibnamefont {Casson}}, \bibinfo
  {author} {\bibfnamefont {D.}~\bibnamefont {Farina}}, \bibinfo {author}
  {\bibfnamefont {L.}~\bibnamefont {Figini}}, \bibinfo {author} {\bibfnamefont
  {T.}~\bibnamefont {Goodman}}, \bibinfo {author} {\bibfnamefont
  {O.}~\bibnamefont {Maj}}, \bibinfo {author} {\bibfnamefont {O.}~\bibnamefont
  {Sauter}}, \bibinfo {author} {\bibfnamefont {H.}~\bibnamefont {Weber}},
  \bibinfo {author} {\bibfnamefont {H.}~\bibnamefont {Zohm}}, \bibinfo {author}
  {\bibfnamefont {G.}~\bibnamefont {Saibene}}, \ and\ \bibinfo {author}
  {\bibfnamefont {M.}~\bibnamefont {Henderson}},\ }\bibfield  {title} {\enquote
  {\bibinfo {title} {On recent results in the modelling of
  neoclassical-tearing-mode stabilization via electron cyclotron current drive
  and their impact on the design of the upper {EC} launcher for {ITER}},}\
  }\href {\doibase 10.1088/0029-5515/55/1/013023} {\bibfield  {journal}
  {\bibinfo  {journal} {Nuclear Fusion}\ }\textbf {\bibinfo {volume} {55}},\
  \bibinfo {pages} {013023} (\bibinfo {year} {2015})}\BibitemShut {NoStop}%
\bibitem [{\citenamefont {Sauter}(2004)}]{Sauter_2004}%
  \BibitemOpen
  \bibfield  {author} {\bibinfo {author} {\bibfnamefont {O.}~\bibnamefont
  {Sauter}},\ }\bibfield  {title} {\enquote {\bibinfo {title} {On the
  contribution of local current density to neoclassical tearing mode
  stabilization},}\ }\href {\doibase 10.1063/1.1787791} {\bibfield  {journal}
  {\bibinfo  {journal} {Physics of Plasmas}\ }\textbf {\bibinfo {volume}
  {11}},\ \bibinfo {pages} {4808--4813} (\bibinfo {year} {2004})}\BibitemShut
  {NoStop}%
\bibitem [{\citenamefont {Zhang}\ \emph {et~al.}(2019)\citenamefont {Zhang},
  \citenamefont {Ma}, \citenamefont {Zhang},\ and\ \citenamefont
  {Zhu}}]{Zhang_1019}%
  \BibitemOpen
  \bibfield  {author} {\bibinfo {author} {\bibfnamefont {W.}~\bibnamefont
  {Zhang}}, \bibinfo {author} {\bibfnamefont {Z.~W.}\ \bibnamefont {Ma}},
  \bibinfo {author} {\bibfnamefont {Y.}~\bibnamefont {Zhang}}, \ and\ \bibinfo
  {author} {\bibfnamefont {J.}~\bibnamefont {Zhu}},\ }\bibfield  {title}
  {\enquote {\bibinfo {title} {Stabilization of tearing modes by modulated
  electron cyclotron current drive},}\ }\href {\doibase 10.1063/1.5080379}
  {\bibfield  {journal} {\bibinfo  {journal} {AIP Advances}\ }\textbf {\bibinfo
  {volume} {9}},\ \bibinfo {pages} {015020} (\bibinfo {year}
  {2019})}\BibitemShut {NoStop}%
\bibitem [{\citenamefont {Yu}, \citenamefont {Zhang},\ and\ \citenamefont
  {Günter}(2004)}]{Yu_2004}%
  \BibitemOpen
  \bibfield  {author} {\bibinfo {author} {\bibfnamefont {Q.}~\bibnamefont
  {Yu}}, \bibinfo {author} {\bibfnamefont {X.~D.}\ \bibnamefont {Zhang}}, \
  and\ \bibinfo {author} {\bibfnamefont {S.}~\bibnamefont {Günter}},\
  }\bibfield  {title} {\enquote {\bibinfo {title} {Numerical studies on the
  stabilization of neoclassical tearing modes by radio frequency current
  drive},}\ }\href {\doibase 10.1063/1.1710521} {\bibfield  {journal} {\bibinfo
   {journal} {Physics of Plasmas}\ }\textbf {\bibinfo {volume} {11}},\ \bibinfo
  {pages} {1960--1968} (\bibinfo {year} {2004})}\BibitemShut {NoStop}%
\bibitem [{\citenamefont {Wang}\ \emph {et~al.}(2015)\citenamefont {Wang},
  \citenamefont {Zhang}, \citenamefont {Wu}, \citenamefont {Zhu},\ and\
  \citenamefont {Hu}}]{Wang_2015}%
  \BibitemOpen
  \bibfield  {author} {\bibinfo {author} {\bibfnamefont {X.}~\bibnamefont
  {Wang}}, \bibinfo {author} {\bibfnamefont {X.}~\bibnamefont {Zhang}},
  \bibinfo {author} {\bibfnamefont {B.}~\bibnamefont {Wu}}, \bibinfo {author}
  {\bibfnamefont {S.}~\bibnamefont {Zhu}}, \ and\ \bibinfo {author}
  {\bibfnamefont {Y.}~\bibnamefont {Hu}},\ }\bibfield  {title} {\enquote
  {\bibinfo {title} {Numerical study on the stabilization of neoclassical
  tearing modes by electron cyclotron current drive},}\ }\href {\doibase
  10.1063/1.4913352} {\bibfield  {journal} {\bibinfo  {journal} {Physics of
  Plasmas}\ }\textbf {\bibinfo {volume} {22}},\ \bibinfo {pages} {022512}
  (\bibinfo {year} {2015})}\BibitemShut {NoStop}%
\bibitem [{\citenamefont {Chen}\ \emph {et~al.}(2015)\citenamefont {Chen},
  \citenamefont {Liu}, \citenamefont {Sun}, \citenamefont {Duan},\ and\
  \citenamefont {Sun}}]{Chen_2015}%
  \BibitemOpen
  \bibfield  {author} {\bibinfo {author} {\bibfnamefont {L.}~\bibnamefont
  {Chen}}, \bibinfo {author} {\bibfnamefont {J.}~\bibnamefont {Liu}}, \bibinfo
  {author} {\bibfnamefont {G.}~\bibnamefont {Sun}}, \bibinfo {author}
  {\bibfnamefont {P.}~\bibnamefont {Duan}}, \ and\ \bibinfo {author}
  {\bibfnamefont {J.}~\bibnamefont {Sun}},\ }\bibfield  {title} {\enquote
  {\bibinfo {title} {Modeling of the influences of electron cyclotron current
  drive on neoclassical tearing modes},}\ }\href {\doibase 10.1063/1.4921669}
  {\bibfield  {journal} {\bibinfo  {journal} {Physics of Plasmas}\ }\textbf
  {\bibinfo {volume} {22}},\ \bibinfo {pages} {052120} (\bibinfo {year}
  {2015})}\BibitemShut {NoStop}%
\bibitem [{\citenamefont {Widmer}\ \emph {et~al.}(2019)\citenamefont {Widmer},
  \citenamefont {Maget}, \citenamefont {F{\'{e}}vrier}, \citenamefont
  {Lütjens},\ and\ \citenamefont {Garbet}}]{Widmer_2019}%
  \BibitemOpen
  \bibfield  {author} {\bibinfo {author} {\bibfnamefont {F.}~\bibnamefont
  {Widmer}}, \bibinfo {author} {\bibfnamefont {P.}~\bibnamefont {Maget}},
  \bibinfo {author} {\bibfnamefont {O.}~\bibnamefont {F{\'{e}}vrier}}, \bibinfo
  {author} {\bibfnamefont {H.}~\bibnamefont {Lütjens}}, \ and\ \bibinfo
  {author} {\bibfnamefont {X.}~\bibnamefont {Garbet}},\ }\bibfield  {title}
  {\enquote {\bibinfo {title} {Non-linear simulations of neoclassical tearing
  mode control by externally driven {RF} current and heating, with application
  to {ITER}},}\ }\href {\doibase 10.1088/1741-4326/ab300f} {\bibfield
  {journal} {\bibinfo  {journal} {Nuclear Fusion}\ }\textbf {\bibinfo {volume}
  {59}},\ \bibinfo {pages} {106012} (\bibinfo {year} {2019})}\BibitemShut
  {NoStop}%
\bibitem [{\citenamefont {Helander}(2012)}]{Helander_2017}%
  \BibitemOpen
  \bibfield  {author} {\bibinfo {author} {\bibfnamefont {P.}~\bibnamefont
  {Helander}},\ }\bibfield  {title} {\enquote {\bibinfo {title} {Classical and
  neoclassical transport in tokamaks},}\ }\href {\doibase
  10.13182/FST12-A13500} {\bibfield  {journal} {\bibinfo  {journal} {Fusion
  Science and Technology}\ }\textbf {\bibinfo {volume} {61}},\ \bibinfo {pages}
  {133--141} (\bibinfo {year} {2012})}\BibitemShut {NoStop}%
\bibitem [{\citenamefont {Konovalov}\ \emph {et~al.}(2005)\citenamefont
  {Konovalov}, \citenamefont {Mikhailovskii}, \citenamefont {Ozeki},
  \citenamefont {Takizuka}, \citenamefont {Shirokov},\ and\ \citenamefont
  {Hayashi}}]{Konovalov_2005}%
  \BibitemOpen
  \bibfield  {author} {\bibinfo {author} {\bibfnamefont {S.~V.}\ \bibnamefont
  {Konovalov}}, \bibinfo {author} {\bibfnamefont {A.~B.}\ \bibnamefont
  {Mikhailovskii}}, \bibinfo {author} {\bibfnamefont {T.}~\bibnamefont
  {Ozeki}}, \bibinfo {author} {\bibfnamefont {T.}~\bibnamefont {Takizuka}},
  \bibinfo {author} {\bibfnamefont {M.~S.}\ \bibnamefont {Shirokov}}, \ and\
  \bibinfo {author} {\bibfnamefont {N.}~\bibnamefont {Hayashi}},\ }\bibfield
  {title} {\enquote {\bibinfo {title} {Role of anomalous transport in onset and
  evolution of neoclassical tearing modes},}\ }\href {\doibase
  10.1088/0741-3335/47/12b/s17} {\bibfield  {journal} {\bibinfo  {journal}
  {Plasma Physics and Controlled Fusion}\ }\textbf {\bibinfo {volume} {47}},\
  \bibinfo {pages} {B223--B236} (\bibinfo {year} {2005})}\BibitemShut {NoStop}%
\bibitem [{\citenamefont {Choi}\ \emph {et~al.}(2018)\citenamefont {Choi},
  \citenamefont {Haye}, \citenamefont {Lanctot}, \citenamefont {Olofsson},
  \citenamefont {Strait}, \citenamefont {Sweeney},\ and\ \citenamefont
  {and}}]{Choi_2018}%
  \BibitemOpen
  \bibfield  {author} {\bibinfo {author} {\bibfnamefont {W.}~\bibnamefont
  {Choi}}, \bibinfo {author} {\bibfnamefont {R.~L.}\ \bibnamefont {Haye}},
  \bibinfo {author} {\bibfnamefont {M.}~\bibnamefont {Lanctot}}, \bibinfo
  {author} {\bibfnamefont {K.}~\bibnamefont {Olofsson}}, \bibinfo {author}
  {\bibfnamefont {E.}~\bibnamefont {Strait}}, \bibinfo {author} {\bibfnamefont
  {R.}~\bibnamefont {Sweeney}}, \ and\ \bibinfo {author} {\bibfnamefont
  {F.~V.}\ \bibnamefont {and}},\ }\bibfield  {title} {\enquote {\bibinfo
  {title} {Feedforward and feedback control of locked mode phase and rotation
  in {DIII}-d with application to modulated {ECCD} experiments},}\ }\href
  {\doibase 10.1088/1741-4326/aaa6e3} {\bibfield  {journal} {\bibinfo
  {journal} {Nuclear Fusion}\ }\textbf {\bibinfo {volume} {58}},\ \bibinfo
  {pages} {036022} (\bibinfo {year} {2018})}\BibitemShut {NoStop}%
\end{thebibliography}%

\end{document}